\documentclass{aa}  
\usepackage[varg]{txfonts}

\usepackage{amsmath}
\usepackage{nccmath}
\usepackage{amssymb}
\usepackage{graphicx}
\usepackage{color}
\usepackage{comment}
\usepackage[normalem]{ulem}
\usepackage{enumitem}
\usepackage{xspace}

\usepackage{xcolor}
\definecolor{xlinkcolor}{cmyk}{1,1,0,0}
\definecolor{Green}{cmyk}{1,0,1,0}
\usepackage[breaklinks=true,    
     colorlinks=true,    
     linkcolor=xlinkcolor,     
     citecolor=xlinkcolor,     
     filecolor=xlinkcolor,  
     urlcolor=xlinkcolor]{hyperref}

\usepackage{pifont}
\newcommand{\tder}[2]{\frac{\mathrm{d}#1}{\mathrm{d}#2}}
\newcommand{\pder}[2]{\frac{\partial #1}{\partial #2}}

\bibpunct{(}{)}{;}{a}{}{,} 

\usepackage[encapsulated]{CJK}
\usepackage{ucs}
\newcommand{\tctext}[1]{\begin{CJK}{UTF8}{bkai}#1\end{CJK}}

\begin{document}

   \title{The impact of dust evolution on the dead zone outer edge in magnetized protoplanetary disks}

   \author{Timmy~N.~Delage\inst{1},
           Mat\'ias~G\'arate\inst{1,2},
           Satoshi~Okuzumi\inst{3},
           Chao-Chin~Yang (\tctext{楊朝欽})\inst{4,5},
           Paola~Pinilla\inst{1,6},
           Mario~Flock\inst{1,7},
           Sebastian~Markus~Stammler\inst{2},
           \and Tilman~Birnstiel\inst{2,8}
           }

   \institute{Max-Planck-Institut f\"{u}r Astronomie, K\"{o}nigstuhl 17, 69117, Heidelberg, Germany, \email{delage@mpia.de}
   \and University Observatory, Faculty of Physics, Ludwig-Maximilians-Universit\"at M\"unchen, Scheinerstr.\ 1, 81679 Munich, Germany
   \and Department of Earth and Planetary Sciences, Tokyo Institute of Technology, Meguro, Tokyo 152-8551, Japan
   \and Department of Physics and Astronomy, University of Nevada, Las Vegas, 4505 South Maryland Parkway, Box~454002, Las Vegas, NV 89154-4002, USA
   \and Department of Physics and Astronomy, The University of Alabama, Box~870324, Tuscaloosa, AL~35487-0324, USA
   \and Mullard Space Science Laboratory, University College London, Holmbury St Mary, Dorking, Surrey RH5 6NT, UK
   \and Jet Propulsion Laboratory, California Institute of Technology, Pasadena, CA 91109, USA
   \and Exzellenzcluster ORIGINS, Boltzmannstr. 2, D-85748 Garching, Germany
   } 

   \date{Received date / 
   Accepted date}

\titlerunning{The impact of dust evolution on the dead zone outer edge in magnetized protoplanetary disks}
\authorrunning{T.~N.~Delage~et~al.~}

 
  \abstract
   {The dead zone outer edge corresponds to the transition from the magnetically dead to the magnetorotational instability (MRI)-active regions in the outer protoplanetary disk mid-plane. It has been previously hypothesized to be a sweet spot for dust particles trapping. A more consistent approach to access such an idea yet remains to be developed, since the interplay between dust evolution and MRI-driven accretion over million years has been poorly understood.}
   {We provide an important step toward a better understanding of the MRI--dust coevolution in protoplanetary disks. In this pilot study, we present a proof of concept that dust evolution ultimately plays a crucial role in the MRI activity.}
   {First, we study how a fixed power-law dust size distribution with varying parameters impacts the MRI activity, especially the steady-state MRI-driven accretion, by employing and improving our previous 1+1D MRI-driven turbulence model. Second, we relax the steady-state accretion assumption in this disk accretion model, and partially couple it to a dust evolution model in order to investigate how the evolution of dust (dynamics and grain growth processes combined) and MRI-driven accretion are intertwined on million-year timescales, from a more sophisticated modeling of the gas ionization degree.}
   {Dust coagulation and settling lead to a higher gas ionization degree in the protoplanetary disk, resulting in stronger MRI-driven turbulence as well as a more compact dead zone. On the other hand, fragmentation has an opposite effect because it replenishes the disk in small dust particles which are very efficient in sweeping up free electrons and ions from the gas phase. Since the dust content of the disk decreases over million years of evolution due to radial drift, the MRI-driven turbulence overall becomes stronger and the dead zone more compact until the disk dust-gas mixture eventually behaves as a grain-free plasma. Furthermore, our results show that dust evolution alone does not lead to a complete reactivation of the dead zone. For typical T-Tauri stars, we find that the dead zone outer edge is expected to be located roughly between $10\,$au and $50\,$au during the disk lifetime for our choice of the magnetic field strength and configuration. Finally, the MRI activity evolution is expected to be crucially sensitive to the choice made for the minimum grain size of the dust distribution.}
   {The MRI activity evolution (hence the temporal evolution of the MRI-induced $\alpha$-parameter) is controlled by dust evolution and occurs on a timescale of local dust growth, as long as there is enough dust particles in the disk to dominate the recombination process for the ionization chemistry. Once it is no longer the case, the MRI activity evolution is expected to be controlled by gas evolution and occurs on a viscous evolution timescale.}

    
   \maketitle
%

\section{Introduction} \label{sect:intro}

With a typical lifetime of a few million years \citep[][]{2001ApJ...553L.153H, 2021arXiv210406838V}, protoplanetary disks are known to rapidly accrete their gas and dust content onto the central pre-main-sequence star, with a typical accretion rate of $\sim 10^{-9}$--$10^{-8} \, M_{\odot}.\rm{yr}^{-1}$ \citep[e.g.,][]{doi:10.1146/annurev-astro-081915-023347}. The accretion phenomenon is ultimately controlled by angular momentum transport and outflow mass loss processes. Such processes shape the disk structure and global evolution as well as its dispersal, hence being of crucial importance in understanding the first steps of planet formation \citep[][]{Armitage_2011, 2019SAAS...45....1A}. Particularly, the disk turbulence shapes the global density distribution within which planets forms, and strongly impacts the evolution of dust particles which are the building blocks of planets \citep[][]{1995Icar..114..237D, Ormel2007, Youdin2007, Birnstiel2010}. Despite its fundamental role, the nature of turbulence in protoplanetary disks is not well constrained yet. Pure hydrodynamic-driven mechanisms have been suggested for generating turbulence such as the vertical shear instability \citep[VSI; e.g.,][]{10.1111/j.1365-8711.1998.01118.x, Nelson_2013, Lin_2015, Manger_2020, 2020ApJ...897..155F}, or the baroclinic instabilities \citep[e.g.,][]{Klahr_2003, Raettig_2013}. Nevertheless, theoretical studies show that the resulting turbulence level is typically too weak to explain the observed accretion rates of protoplanetary disks \citep[see the review of][]{2022arXiv220309821L}. Other mechanisms such as the gravitational instability \citep[GI; e.g.,][]{10.1093/mnras/225.3.607, Lodato_2004, Vorobyov_2009} can provide significant turbulence, but only in the early stages of the disk evolution. Consequently, magnetohydrodynamic (MHD)-driven mechanisms such as MHD winds \citep[e.g.,][]{10.1093/mnras/199.4.883, Suzuki_2009, Bai_2016a, Bai_2016b} and the magnetorotational instability \citep[MRI; e.g.,][]{1991ApJ...376..214B, 1998RvMP...70....1B, 1995ApJ...440..742H} are currently thought to be the main candidates for driving disk accretion.  

The MRI-driven accretion can be substantially modified and even suppressed at some locations in the disk by three nonideal MHD effects: Ohmic resistivity, the Hall effect, and ambipolar diffusion. They arise due to the weak level of ionization in the disk \citep[e.g.,][]{1996ApJ...457..355G, 2000ApJ...530..464F, 2002ApJ...570..314S, 2002ApJ...577..534S, 2003ApJ...585..908F, 2005ApJ...628L.155I,2006A&A...445..223I, 2007ApJ...659..729T, 2010ApJ...708..188T, 2008ApJ...679L.131T, 2011ApJ...735....8P} or because of the high drift between the ions and electrons in regions of low gas densities and strong magnetic field strengths \citep[e.g.,][]{2011ApJ...736..144B}. In general, how much and where the MRI activity is suppressed is a complex problem that significantly depends on the dust and gas properties, the magnetic field strength, as well as the complex ionization chemistry \citep[e.g.,][]{2009ApJ...701..737B,2011ApJ...739...50B,Delage2021}. The direct consequence is that a magnetically dead zone arises, characterized by a low gas accretion rate \citep[e.g.,][]{2010A&A...515A..70D}. It causes a steep increase in the disk turbulence at the transition from the dead zone to the MRI-active region, the so-called "dead zone outer edge". This location has been previously hypothesized to be a sweet spot for dust particles trapping, hence potentially explaining some of the observed substructures in protoplanetary disks \citep[e.g.,][]{2012MNRAS.419.1701R,2015A&A...574A..68F, 2016A&A...596A..81P}. 

To further investigate the potential dust trapping power of the dead zone outer edge, one needs to build a model whose outputs can be compared to current dust continuum and gas observations of million-year old disks. Such a model thus necessarily requires a time-dependent framework where nonideal MHD calculations are self-consistently combined with gas and dust evolution on million-year timescales. Some theoretical works study the detailed behavior of MRI-active and non-active regions by performing 3D local shearing box or global simulations \citep[e.g.,][]{2010A&A...515A..70D, 2010ApJ...708..188T, 2011ApJ...736..144B, 2011ApJ...739...50B, 2011ApJ...742...65O, 2011ApJ...735..122F, 2015A&A...574A..68F}. However, these studies did not implement a full treatment for dust evolution (dynamics and grain growth processes combined), and cannot evolve the protoplanetary disk over million years. Conversely, some papers used the Shakura-Sunyaev $\alpha$-disk model \citep{1973A&A....24..337S}, wherein the quantity $\alpha$ encodes the disk turbulence level, in order to make possible on million-year timescales the implementation of gas and dust evolution altogether with an educated guess for the MRI-induced $\alpha$-parameter \citep[e.g.,][]{2016A&A...596A..81P}. Crucially, though, ad hoc prescriptions of $\alpha$ have been adopted, without accounting for the detailed physics of the MRI. 

A more consistent approach to access the dead zone outer edge as a potential location for dust trapping thus yet remains to be developed, since the interplay between dust evolution and MRI-driven accretion has been poorly understood. Particularly, it is still unclear how dust evolution impacts the MRI-driven accretion in protoplanetary disks on million-year timescales. Indeed, most previous works have not implemented the feedback of dust evolution on the ionization state of the disk, which is crucial to accurately describe the MRI activity. A possible way to investigate such interplay is by using a "trade-off" model that combines a 1D viscous disk model and nonideal MHD calculations: the viscosity parameter $\alpha$ is determined by the MRI-driven turbulence accounting for the nonideal MHD effects, with a careful modeling of the gas ionization degree. Such a model allows to capture the essence of the MRI in a non-computationally expensive way, which makes the coupling with 1D gas and dust evolution models (including growth, fragmentation, settling, and radial drift) feasible on million-year timescales. 

\citet{2012ApJ...753L...8O} investigated the impact of dust evolution on the MRI-driven turbulence by coupling the dust coagulation equation, an analytic model for the disk ionization including charged grains, and an empirical model for $\alpha$ based on nonideal MHD simulations. Nevertheless, their work modeled the dust as a two-population phase instead of considering the full dust size distribution, neglected dust radial drift, and most importantly did not include ambipolar diffusion which is fundamental to accurately describe the MRI activity in the outer part of protoplanetary disks. 

\citet{Delage2021} put forward a trade-off model specifically designed to study the MRI-driven accretion in the outer protoplanetary disk ($r \gtrsim 1\,$au, where $r$ is the distance from the central star). Their model allowed to compute a self-consistent MRI-induced $\alpha$-parameter given stellar, gas and dust properties, in the framework of viscously driven accretion, accounting for both Ohmic resistivity and ambipolar diffusion with a careful modeling of the gas ionization degree. In their paper, they provided some insights regarding the potential impact of gas and stellar evolution on the MRI activity, particularly the dead zone outer edge. They showed that the MRI-driven turbulence becomes stronger when the total disk gas mass decreases due to a higher gas ionization degree. Additionally, they indirectly showed that a higher stellar X-ray luminosity leads to stronger MRI-driven turbulence overall due to a higher stellar X-ray ionization rate. However, their study assumed a fixed grain size across the whole protoplanetary disk, and did not investigate the interplay between dust evolution and MRI-driven accretion over million years. 

The aim of this present paper is thus twofold: (1) Understand how the implementation of a dust size distribution impacts the MRI activity, especially the steady-state MRI-driven accretion described in \citet{Delage2021}; (2) Present a pilot study providing a proof of concept that dust evolution alone has a substantial impact on the MRI-driven turbulence, particularly the dead zone outer edge. Such a study solely focusing on the effect of dust evolution is a necessary first step toward a better understanding of the MRI--dust coevolution. Here we note that the accretion driven by magnetic disk winds is ignored to focus our efforts on the MRI. 

The layout of the paper is as follows. In Sect.~\ref{sect:model} we present the disk model employed. In Sect.~\ref{sect:Setup} we describe the numerical implementation of the various simulations making use of our disk model. In Sect.~\ref{sect:results} we present the results that investigate the impact of a dust size distribution as well as dust evolution on the MRI activity. In Sect.~\ref{sect:discussion} we discuss the implications of our results. Finally, Sect.~\ref{sect:summary and conclusions} summarizes our conclusions.

\section{Disk Model} \label{sect:model}

We consider a central star of mass $M_\star$ and bolometric luminosity $L_\star$. Additionally, we assume that the envelope has dispersed to reveal a gravitationally stable and viscously accreting disk of total disk gas mass $M_{\rm{disk}}$. The protoplanetary disk is considered geometrically thin, so that the vertical and radial dimensions can be decoupled into a 1+1D $(r,z)$ framework, where each radial grid-point contains an independent vertical grid. Furthermore, we assume the disk to be axisymmetric and symmetric about the mid-plane.

The local mass and angular momentum transport are assumed to be solely controlled by the MRI and hydrodynamic instabilities based on the model of \citet{Delage2021}, where the disk turbulence level is encoded into the viscosity parameter $\alpha$. In their 1+1D model ($r-z$ plane), the solution for vertical stratification of gas and dust is required in order to get an approximate 1D description of vertically layered accretion within the Shakura-Sunyaev viscous $\alpha$-disk model. The main output of their model is thus the effective turbulent parameter $\bar{\alpha}$, defined as the pressure-weighted vertical average of the local turbulent parameter $\alpha$:
\begin{equation}
    \bar{\alpha}(r) = \frac{\int_{-\infty}^{+\infty} \alpha(r,z) \: P_{\rm{gas}}(r,z) \: dz}{\int_{-\infty}^{+\infty} P_{\rm{gas}}(r,z) \: dz},
    \label{eq:alpha bar}
\end{equation}

\noindent where $z$ is the height from the mid-plane, and $P_{\rm{gas}}$ is the isothermal gas pressure. 

In Sects.~\ref{sect:gas properties} and \ref{sect:dust properties}, we present the disk properties for the gas and the dust, respectively. From these, we can obtain the effective turbulent parameter, $\bar{\alpha}$, by employing the MRI-driven turbulence model described in Sect.~\ref{sect:MRI-driven turbulence}.

\subsection{Gas} \label{sect:gas properties}

The gas is assumed vertically isothermal, with a radial temperature profile set by passively absorbing stellar irradiation
\begin{equation}
T(r) = \left[T^{4}_{\rm{1 \,au}} \left(\frac{r}{1 \, \rm{au}}\right)^{-2} \left(\frac{L_\star}{L_{\odot}}\right) + T^{4}_{\rm{bkg}}\right]^{\frac{1}{4}},
\label{eq:gas temperature}
\end{equation}

\noindent where $T_{\rm{1 \,au}} = 280 \,$K is the gas temperature at $r = 1 \,$au for $L_\star = L_{\odot}$, and $T_{\rm{bkg}} = 10 \,$K is the background gas temperature corresponding to the primordial temperature of the cloud prior to the collapse. We note that the choice made for the gas temperature model does not have a significant impact on the MRI-driven turbulence, since its temperature dependence is weak \citep[see Appendix~D.3 of][]{Delage2021}.

Assuming hydrostatic equilibrium in the vertical direction gives the gas volume density profile
\begin{equation}
    \rho_{\rm{gas}}(r,z) = \frac{\Sigma_{\rm{gas}}(r)}{\sqrt{2 \pi} H_{\rm{gas}}(r)} \exp{\left(-\frac{z^2}{2 H^{2}_{\rm{gas}}(r)}\right)},
    \label{eq:volume gas density}
\end{equation}

\noindent where $\Sigma_{\rm{gas}}$ is the gas surface density chosen as explained in Sect.~\ref{sect:Setup}, and $H_{\rm{gas}} = c_s/\Omega_K$ is the disk gas scale height with the isothermal sound speed $c_s = \sqrt{k_B T/\mu m_{\rm H}}$ and the Keplerian angular velocity $\Omega_K = \sqrt{G M_\star/r^3}$. Here $k_B$ is the Boltzmann constant, $\mu = 2.34$ is the mean molecular weight (assuming solar abundances), $m_{\rm H}$ is the atomic mass of hydrogen, and $G$ is the gravitational constant. 

Finally, the total number density of gas particles is defined as $n_{\rm{gas}} = \rho_{\rm{gas}}/m_{\rm{neutral}}$, with $m_{\rm{neutral}} = \mu m_{\rm H}$ the mean molecular mass.

\subsection{Dust particles} \label{sect:dust properties}

Observations of protoplanetary disks at different wavelengths suggest that the dust particles can significantly decouple from the gas, depending on the grain size \citep[see e.g.][]{Andrews2016, vanBoekel2017, Huang2018}. Consequently, including a dust model is required.

We considered that the dust phase consists of a distribution of dust particles with different sizes. Each grain is assumed to be a perfect compact sphere of intrinsic volume density $\rho_{\rm{bulk}} = 1.4 \,$g.cm$^{-3}$ \citep[consistent with the solar abundance when H$_2$O ice is included in grains; see][]{1994ApJ...421..615P}. For a grain of size $a$, the corresponding mass is then $m(a) = \frac{4}{3} \pi \rho_{\rm{bulk}} a^{3}$.

Most of the transport and dynamics of dust particles in protoplanetary disks are regulated by the interactions between themselves and the gas. A way to quantify the importance of the drag forces on the dynamics of a dust particle (hence the level of coupling between that dust particle and the gas) is by its Stokes number, defined as the dimensionless version of the stopping time of that particle. Near the mid-plane, the Stokes number of a dust particle of size $a$ is given by
\begin{equation}
    {\rm St}(r,a) =  \frac{\pi}{2}\frac{a\, \rho_{\rm{bulk}}}{\Sigma_{\rm{gas}}(r)} \cdot
    \begin{cases}
				1 & \lambda_{\rm{mfp}}/a \geq 4/9\\
                \frac{4}{9} \frac{a}{\lambda_{\rm{mfp}}} & \lambda_{\rm{mfp}}/a < 4/9,
	\end{cases} 
	\label{eq:stokes number}
\end{equation}

\noindent where $\lambda_{\rm{mfp}} = (n_{\rm{gas}} \sigma_{\rm{H}_2})^{-1}$ is the mean free path of gas particles, with $\sigma_{\rm{H}_2} = 2 \times 10^{-15} \,$cm$^2$ the molecular cross-section for H$_2$ \citep[e.g.,][]{2008A&A...487L...1B,Birnstiel2010}.

\subsubsection{Dust settling} \label{sect:dust settling}

Dust models predict that grains tend to settle more efficiently toward the mid-plane as they grow in size \citep{1995Icar..114..237D}. Additionally, dust settling appears to be at play by ALMA observations of edge-on disks \citep[see, e.g.,][]{2019A&A...624A...7V, 2020A&A...642A.164V}. As a result, we can expect the number of dust particles to drop significantly above a dust scale height, $H_{\rm{dust}}$, that can be much smaller than the gas scale height, $H_{\rm{gas}}$. Assuming dust stirring to be induced by the MRI-driven turbulence, we can relate $H_{\rm{dust}}$, $H_{\rm{gas}}$, St, and $\bar{\alpha}$ by \citep[][]{1995Icar..114..237D,Youdin2007,2018ApJ...868...27Y} 
\begin{equation}
    H_{\rm{dust}}(r,a) = H_{\rm{gas}}(r) \sqrt{\frac{\bar{\alpha}(r)}{\bar{\alpha}(r) + \rm{St}(r,a)}}. 
    \label{eq:dust scale height}
\end{equation}

\noindent This expression is given for each grain species of size $a$. We assume that ${\rm St}/D_{\rm{gas}}$ is independent of $z$, where $D_{\rm{gas}}$ is the gas diffusion coefficient. We also implicitly approximate $D_{\rm{gas}}$ by the vertically integrated gas kinematic viscosity 
\begin{equation}
    \bar{\nu} = \bar{\alpha} c_s H_{\rm gas}.
\end{equation}

In the vertical direction, we then assume that the dust volume density profile follows a Gaussian distribution. For each grain species of size $a$, it is described by 
\begin{equation}
    \rho_{\rm{dust}}(r,z,a) = \frac{\Sigma_{\rm{dust}}(r,a)}{\sqrt{2 \pi} H_{\rm{dust}}(r,a)} \exp{\left(-\frac{z^2}{2 H^{2}_{\rm{dust}}(r,a)}\right)}
    \label{eq:volume dust density}
\end{equation}

\noindent where $\Sigma_{\rm{dust}}$ is the dust surface density for each grain species of size $a$ that we describe in Sect.~\ref{sect:dust surface density}. The total dust volume density (accounting for all grain species) is thus defined as 
\begin{equation}
    \rho_{\rm{dust,\,tot}}(r,z) = \sum_{a}^{} \: \rho_{\rm{dust}}(r,z,a).
    \label{eq:total dust density}
\end{equation}

Finally, the number density for each grain species of size $a$ is given by
\begin{equation}
    n_{\rm{dust}}(r,z,a) = \frac{\rho_{\rm{dust}}(r,z,a)}{m(a)},
\end{equation}
\noindent with $m(a)$ the corresponding grain mass. It follows that the total number density (accounting for all grain species) is determined by
\begin{equation}
    n_{\rm{dust,\,tot}}(r,z) = \sum_{a}^{} \: n_{\rm{dust}}(r,z,a).
    \label{eq:total dust number density}
\end{equation}

\subsubsection{Dust surface density} \label{sect:dust surface density}

The remaining quantity to complete the description of the dust phase is the surface density for each grain species of size $a$ ($\Sigma_{\rm{dust}}(r,a)$). In this paper, it is determined either by assuming that the dust size distribution follows a fixed power-law, or is directly obtained from a dust evolution model.

\paragraph{Power-law dust size distribution.} 

It is determined by the three parameters $a_{\rm{min}}$, $a_{\rm{dist,Max}}$, and $p_{\rm{dist,Exp}}$; respectively the distribution minimum grain size, maximum grain size, and exponent. In the grain size range $\left[ a, a + da \right]$, it reads:
\begin{equation}
    n'_{\rm{dust}}(a) \, da \propto \left\{
                    \begin{array}{ll}
                         a^{p_{\rm{dist,Exp}}} \, da & \mbox{if} \, a_{\rm{min}} \leq a \leq a_{\rm{dist,Max}}  \\
                         0 & \mbox{otherwise}
                    \end{array}
                \right.
    \label{eq:power-law dust distribution}
\end{equation}

\noindent where $n'_{\rm{dust}}(a)$ refers to the dust number density per grain size, and is a distribution function over $a$. It differs from $n_{\rm{dust}}(a)$ above, which is already the quantity integrated over the bin size around $a$.

$\Sigma_{\rm{dust}}(r,a)$ then follows from the conservation of the total dust mass: The quantity $\Sigma_{\rm{dust,tot}}(r) = \sum_{a}^{} \: \Sigma_{\rm{dust}}(r,a)$ must be equal to $f_{\rm{dg,tot}}(r) \Sigma_{\rm{gas}}(r)$, where $f_{\rm{dg,tot}}(r)$ is the vertically integrated total dust-to-gas mass ratio (accounting for all grain species) at radius $r$. We note that $f_{\rm{dg,tot}}$ is a free-parameter for the models of this present paper using the power-law dust size distribution (Models I--III). We assume this quantity to be radially constant equal to $10^{-2}$. 

\paragraph{Dust evolution model.} 

The dust size distribution can substantially differ from a power-law because dust particles collide with each other leading to coagulation or fragmentation, and are transported across the disk by different mechanisms such as thermal Brownian motion, vertical stirring and settling, turbulent mixing, and drift (azimuthal and radial). As a result, we need a dust evolution model that includes all these mechanisms. For this purpose, we used the code \texttt{DustPy}\footnote{\href{https://github.com/stammler/DustPy}{github.com/stammler/DustPy}.} \citep[][]{2022arXiv220700322S}, which can simulate the advection of gas and dust, along with the growth and fragmentation of multiple grain species, based on the model of \citet{Birnstiel2010}. Below, we summarize the main ingredients in the model.

Following \citet{Nakagawa1986}  and \citet{Takeuchi2002}, the dust radial velocity for each grain species of size $a$ is given by:
\begin{equation} 
    v_{\rm{r,dust}}(r,a) = \frac{v_{\rm{gas}}(r)}{1 + \rm{St}^2(r,a)} -  \frac{2 \rm{St}(r,a)}{1 + \rm{St}^2(r,a)} \eta(r) v_K(r),
    \label{eq:dust_radial_velocity}
\end{equation}

\noindent with $\eta = -\frac{1}{2} \, (H_{\rm{gas}} / r)^2\, (d\ln{P_{\rm{gas,mid}}}/d\ln{r})$ the gas pressure support parameter \citep[e.g.,][]{2016SSRv..205...41B}, $v_K = r \Omega_K$ the Keplerian orbital velocity, and $P_{\rm{gas,mid}}$ the mid-plane isothermal gas pressure $P_{\rm{gas}} = \rho_{\rm{gas}} c_s^{2}$. Moreover, $v_{\rm{gas}}$ corresponds to the gas viscous velocity, and is determined by
\begin{equation}
    v_{\rm{gas}} = - \frac{3}{\Sigma_{\rm{gas}} \sqrt{r}} \frac{\partial}{\partial r}\left( \Sigma_{\rm{gas}} \bar{\nu} \sqrt{r}\right),
\end{equation}

In addition to advection, each dust particle of size $a$ diffuses according to the concentration gradient, with a dust radial diffusivity approximated by \citep{Youdin2007} 
\begin{equation}
    D_{\rm{dust}}(r,a) = \frac{\bar{\nu}(r)}{1 + \rm{St}^2(r,a)}.
    \label{eq:dust_radial_diffusion_coefficient}
\end{equation}

\noindent Given that both the dust radial and vertical diffusion coefficients are determined by Eq.~\eqref{eq:dust_radial_diffusion_coefficient}, where the latter results in Eq.~\eqref{eq:dust scale height}, we implicitly assume the turbulence to be isotropic. 

The transport for each grain of size $a$ is described by the following 1D radial advection-diffusion equation, for the dust surface density $\Sigma_{\rm{dust}}(r,a)$ \citep{Birnstiel2010}:
\begin{equation} 
 \frac{\partial \Sigma_{\rm{dust}}}{\partial t} + \frac{1}{r} \frac{\partial}{\partial r}\Bigg\{ r \times \left[ \Sigma_{\rm{dust}} v_{\rm{r,dust}} - \Sigma_{\rm{gas}} D_{\rm{dust}} \frac{\partial}{\partial r} \left( \frac{\Sigma_{\rm{dust}}}{\Sigma_{\rm{gas}}} \right) \right]\Bigg\} = 0
 \label{eq:dust_advection_diffusion}
\end{equation}

\noindent There are no sink terms included in Eq.~\eqref{eq:dust_advection_diffusion} because we ignore the potential loss of dust particles through wind entrainment from, for example, internal photoevaporative winds or MHD disk winds \citep[see e.g.,][]{2021A&A...655A..18G,2022A&A...659A..42R}.  

In order for the dust evolution model to be complete, grain coagulation (growth) and fragmentation needs to be included. Indeed, each grain of the dust distribution can be transported across the protoplanetary disk, and their size can also evolve through sticking and fragmentation \citep[][]{Birnstiel2010}. All of these processes are included in \texttt{DustPy} by solving the Smoluchowski coagulation equation \citep[][]{1916ZPhy...17..557S}, simultaneously with the transport of the grains (Eq.~\eqref{eq:dust_advection_diffusion}).

We note that all dust-related quantities (and $\bar{\alpha}$; see Sect.~\ref{sect:MRI-driven turbulence}) become time-dependent when a dust evolution model is employed. Furthermore, $f_{\rm{dg,tot}}(r) = \Sigma_{\rm{dust,tot}}(r)/\Sigma_{\rm{gas}}(r)$ is no longer a free-parameter for the models of this present paper using the dust evolution model (Models IV--VI). Instead, it is determined by the dust evolution calculations, through solving for $\Sigma_{\rm{dust}}(r,a)$ at each time step. The initial $f_{\rm{dg,tot}}$ (required to start the dust evolution model) is chosen to be radially constant equal to $10^{-2}$.

\subsection{MRI-driven turbulence model} \label{sect:MRI-driven turbulence}

\begin{figure*}
\centering
\includegraphics[width=\textwidth]{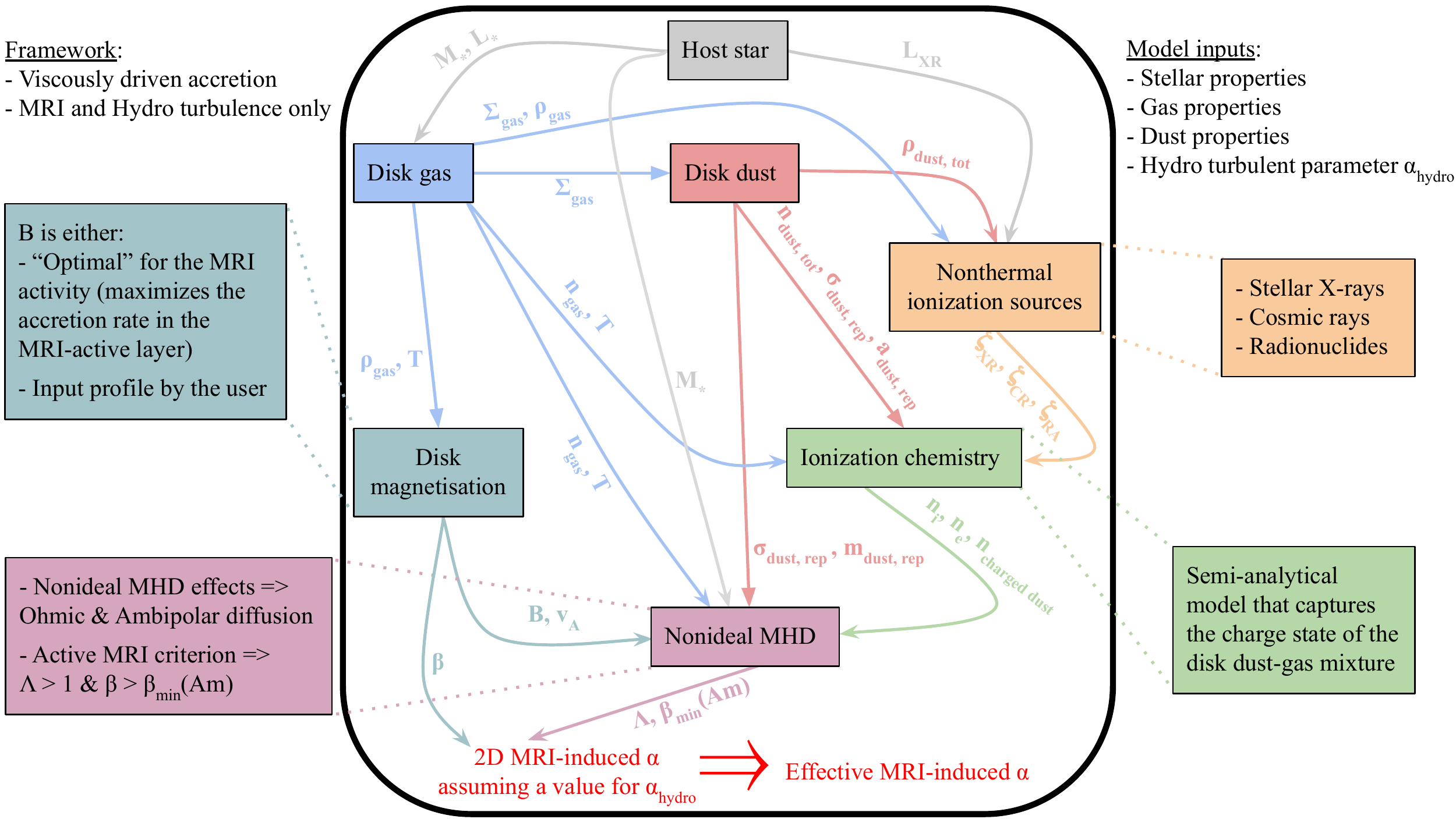}
  \caption{Flowchart of the MRI-driven disk accretion model presented in \citet{Delage2021}. This model captures the essence of the MRI-driven turbulence in a 1+1D framework, accounting for the following: stellar properties (gray symbols), disk gas properties (blue symbols), disk dust properties (red symbols), nonthermal ionization sources (yellow symbols), ionization chemistry modeling the gas ionization degree (green symbols), disk magnetisation properties (powder blue symbols), and nonideal MHD calculations (dark pink symbols). The main output of the model is an effective radial profile for the MRI-induced viscosity parameter, $\bar{\alpha}$ (Eq.~\eqref{eq:alpha bar}). In this paper, we improve the dust phase modeling with a dust size distribution, either by assuming a fixed power-law distribution with different properties or the outputs from dust evolution obtained with \texttt{DustPy} (see Table~\ref{tab:summary models} and text in Sect.~\ref{sect:MRI-driven turbulence} for further explanations).}
 \label{fig:Flowchart_MRI_model}
\end{figure*}

In this paper, we use and improve the MRI-driven turbulence model of \citet{Delage2021}. The main output of this model is an effective radial profile for the MRI-induced viscosity parameter, $\bar{\alpha}$, as shown by the flowchart in Fig.~\ref{fig:Flowchart_MRI_model}. The next paragraph summarizes the main ingredients.  

Their model is a 1+1D global magnetically driven disk accretion model built to study the outer region of class II protoplanetary disks ($r \gtrsim 1\,$au), which accretes viscously solely due to the MRI and hydrodynamic instabilities. It has the advantage to capture the essence of the MRI-driven accretion, without resorting to computationally expensive 3D global nonideal MHD simulations. It includes the key following physical processes: (1) disk heating by passively absorbing stellar irradiation; (2) dust settling; (3) nonthermal ionization from stellar X-rays and galactic cosmic rays as well as the decay of short- and long-lived radionuclides; and (4) ionization chemistry based on a semi-analytical chemical model that captures the charge state of the disk dust-gas mixture, hence carefully modeling the gas ionization degree. In order to know where the MRI can operate in the disk, the general methodology is to compute the magnetic diffusivities of the nonideal MHD effects as well as their corresponding Elsasser numbers from the ionization chemistry, and apply a set of conditions for sustaining active MRI derived from 3D numerical simulations. These conditions account for the suppression of the MRI by Ohmic resistivity and ambipolar diffusion, but ignore for now the role of the Hall effect. In the MRI-dead zones (where the MRI is suppressed), it is further assumed that the gas can still accrete due to a small constant hydrodynamic turbulent parameter $\alpha_{\rm{hydro}}$, induced by hydrodynamic instabilities, such as the VSI \citep[e.g.,][]{2020ApJ...897..155F, 2021arXiv210601159B}. For given stellar, gas and dust properties, the Shakura-Sunyaev viscosity parameter, $\alpha$, can thus be determined self-consistently under the framework of viscously driven accretion from detailed considerations of the MRI with Ohmic resistivity and ambipolar diffusion. It is computed both as a function of radius and height, eventually leading to the effective turbulent parameter $\bar{\alpha}$, which is the key output quantity for further coupling with 1D gas and dust evolution models. In this work, the r.m.s. magnetic field strength ($B$) is numerically constrained by our MRI-driven turbulence model, and chosen such that the MRI activity is at the maximal efficiency as permitted by the two nonideal MHD effects considered. In other words, $B$ is found such that it maximizes the accretion rate in the MRI-active region at any radii \citep[Sect.~3.2 of][]{Delage2021}. We note that the accretion driven by magnetic disk winds is ignored to solely focus on the roles of the MRI and hydrodynamic instabilities. 

In \citet{Delage2021}, the authors assumed a mono-disperse dust distribution of fixed size to describe the dust phase. To improve their model with a dust size distribution, we now define the following three quantities, at any locations $(r,z)$ in the protoplanetary disk: the representative grain size $a_{\rm{dust,rep}}$, the representative grain cross-section $\sigma_{\rm{dust,rep}}$, and the representative grain mass $m_{\rm{dust,rep}}$. They read:
\begin{equation}
    a_{\rm{dust,rep}}(r,z) = \frac{1}{n_{\rm{dust,tot}}} \sum_{a}^{} \: a \: n_{\rm{dust}}(r,z,a),
    \label{eq:representative grain size}
\end{equation}

\begin{equation}
    \sigma_{\rm{dust,rep}}(r,z) = \frac{1}{n_{\rm{dust,tot}}} \sum_{a}^{} \: \pi \: a^{2} n_{\rm{dust}}(r,z,a),
    \label{eq:representative grain cross-section}
\end{equation}
and
\begin{equation}
    m_{\rm{dust,rep}}(r,z) = \frac{1}{\rho_{\rm{dust,tot}}} \sum_{a}^{} \: \frac{4}{3} \: \pi \: \rho_{\rm{bulk}} \: a^{3} \rho_{\rm{dust}}(r,z,a).
    \label{eq:representative grain mass}
\end{equation}

\noindent In practice, $\rho_{\rm{dust}}$, $n_{\rm{dust}}$, $a_{\rm{dust}}$, $\sigma_{\rm{dust}}$, $m_{\rm{dust}}$ defined in \citet{Delage2021} must now be replaced by $\rho_{\rm{dust,tot}}$, $n_{\rm{dust,tot}}$, $a_{\rm{dust,rep}}$, $\sigma_{\rm{dust,rep}}$, $m_{\rm{dust,rep}}$, respectively. Indeed, these new five dust quantities encapsulate all the necessary information to successfully implement a dust size distribution in their model, hence making possible the coupling between dust evolution and MRI-driven turbulence calculations. We note that the definition of $m_{\rm{dust,rep}}$ does not really matter because it only intervenes through the dust Hall parameter, which is actually quasi-independent of it since the grain mass is always much larger than $m_{\rm{neutral}}$ for any grains of size $a \geq 0.1 \, \mu$m \citep[see Eq.~(21) of][]{2007Ap&SS.311...35W}. Furthermore, we note that the representative grain size can be thought of the total grain size per unit volume called "$C_{\rm{tot}}$" in \citet{2011ApJ...731...95O, 2011ApJ...731...96O, 2013ApJ...771...44O}, whereas the representative grain cross-section can be seen as their total grain surface area called "$A_{\rm{tot}}$". We can write $a_{\rm{dust,rep}} = C_{\rm{tot}}/n_{\rm{dust, \: tot}}$ and $\sigma_{\rm{dust,rep}} = A_{\rm{tot}}/n_{\rm{dust, \: tot}}$. For the ionization chemistry, what actually matters are the quantities $C_{\rm{tot}}$ and $A_{\rm{tot}}$. We note that the dust mainly affects the ionization chemistry (hence the MRI-driven turbulence) by $A_{\rm{tot}}$, and weakly by $C_{\rm{tot}}$.

Except for the grain size (here we assume a dust distribution with different sizes rather than a mono-disperse distribution) and the vertically integrated total dust-to-gas-mass ratio $f_{\rm{dg,tot}}$ (see Sect.~\ref{sect:dust surface density}), the parameters used to run our MRI-driven turbulence model are either taken from Table~1 of \citet{Delage2021}, if not explicitly mentioned, or as follows: $M_{\star} = 1 \,$M$_{\odot}$, $L_{\star} = 2 \,$L$_{\odot}$, $M_{\rm{disk}} = 0.05 \,$M$_{\star}$, $\alpha_{\rm{hydro}} = 10^{-4}$, and $s_e = 0.6$. We note that the electron sticking coefficient, $s_e$, is chosen equal to $0.6$ (instead of $0.3$ as in their paper). This updated value is more compatible with the detailed calculations conducted by \citet{2011ApJ...739...50B}.

\section{Simulation setups} \label{sect:Setup}

Table~\ref{tab:summary models} summarizes all the models considered in the present paper. Below we describe them in details, and explain the numerical implementation.
\begin{table*}
    \begin{center}
    \caption{Summary of Models I--VI.}
    \begin{tabular}{ c||c|c|c|c|c|c }
    \hline
    \hline
    Model & $\Sigma_{\rm{gas}}$  & $\Sigma_{\rm{dust}}$ & $a_{\rm{min}}$ & $a_{\rm{dist,Max}}$  & $p_{\rm{dist,Exp}}$ & $v_{\rm{frag}}$ \\
    & [g.cm$^{-2}$] & [g.cm$^{-2}$] & [$\mu$m] & [$\mu$m] & & [m.s$^{-1}$] \\ 
    \hline
    I & SS1 & Power-law  & $0.1$ & $1, 10, 10^{2}, 10^{3} , 10^{4}$ & $-3.5$ & --  \\
 
    II & SS1 & Power-law & $0.1, 0.2, 0.3, 0.4, 0.55$ & $1$ & $-3.5$ & -- \\
   
    III & SS1 & Power-law & $0.1$ & $1$ & $-4.5, -3.5, -2.5, 0.25$ & -- \\
  
    IV & SS2 & \texttt{DustPy} & $0.55$ & -- & -- & $1$ \\
   
    V & LBP  & \texttt{DustPy} & $0.55$ & -- & -- & $1$ \\
    
    VI & LBP & \texttt{DustPy} & $0.55$ & -- & -- & $10$ \\
    \hline
    \hline
\end{tabular}
\tablefoot{For Models I--III, the dust size distribution (hence $\Sigma_{\rm{dust}}$) is based on a power-law defined by the minimum grain size $a_{\rm{min}}$, maximum grain size $a_{\rm{dist,Max}}$, and exponent $p_{\rm{dist,Exp}}$. For Models IV--VI, the dust size distribution is directly obtained from the dust evolution model encoded into the code \texttt{DustPy}, which depends on $a_{\rm{min}}$ and the fragmentation velocity $v_{\rm{frag}}$. For a detailed description of the various models, see Sect.~\ref{sect:Setup}. }
\label{tab:summary models}
\end{center}
\end{table*}

\subsection{Models I--III} \label{sect:Models I--III}

Models I--III investigate how a fixed power-law dust size distribution with varying parameters impacts the MRI activity, especially the steady-state MRI-driven accretion described in \citet{Delage2021}. It is determined by the three parameters $a_{\rm{min}}$, $a_{\rm{dist,Max}}$, and $p_{\rm{dist,Exp}}$ (see Eq.~\eqref{eq:power-law dust distribution}), and its effect is studied by varying them one at a time. We refer to such a dust distribution as "Power-law" in Table~\ref{tab:summary models}. Here we note that providing the fragmentation velocity, $v_{\rm{frag}}$, is irrelevant because there is no dust evolution in this case.

For these models, $\Sigma_{\rm{gas}}$ is computed alongside $\bar{\alpha}$ using our MRI-driven turbulence model (Sect.~\ref{sect:MRI-driven turbulence}) in the following mode: We seek for a steady-state MRI-driven accretion for the gas (the gas accretion rate is radially constant, implying that all regions of the disk accrete at the same rate), corresponding to the given dust and stellar properties. Through an iterative process (see Sect.~4.2 of \citet{Delage2021}), $\bar{\alpha}$ and $\Sigma_{\rm{gas}}$ are computed together in order for the gas to follow such a regime. We refer to it as "Steady-State 1" (SS1) in Table~\ref{tab:summary models}.

\subsection{Models IV--VI} \label{sect:Models IV--VI}

Models IV--VI investigate the impact of dust evolution on the MRI-driven turbulence (there is no gas evolution here).

We particularly monitor how the effective MRI-induced turbulent parameter, $\bar{\alpha}$, varies as a function of various dust evolution snapshots. To do so we partially couple the dust evolution model employed with \texttt{DustPy} to our MRI-driven turbulence model. We emphasize that this coupling is not self-consistent yet, since we have not treated the evolution of dust and $\bar{\alpha}$ simultaneously: At each dust evolution snapshot, the new radial profile of $\bar{\alpha}$ derived from our MRI-driven turbulence model is not re-injected into the dust evolution model. This implies that the feedback from a change in the turbulence level due to dust evolution is not accounted for for the next steps of the dust evolution calculations (although, a change in $\bar{\alpha}$ is expected to impact both the gas and the dust). We justify our choice by reminding that the framework of this present work aims to isolate the effect of dust evolution on the MRI-driven accretion. In this context, our goal is only to investigate whether dust evolution can change the MRI-induced $\bar{\alpha}$ over time at all, as well as where in the protoplanetary disk substantial changes can occur, if any. Besides, we note that a full self-consistent coupling between MRI-driven turbulence calculations, gas and dust evolution will be addressed in a future paper focusing on the potential dust trapping power of the dead zone outer edge.

In this paper, the methodology for partially coupling \texttt{DustPy} with our MRI-driven turbulence model is the following: At each disk radius, we assumed the grain size distribution to initially follow a MRN-like distribution of interstellar grains \citep[][]{1977ApJ...217..425M}, with $a_{\rm{min}} = 0.55 \, \mu$m, $a_{\rm{dist,Max}} = 1 \, \mu$m and $p_{\rm{dist,Exp}} = -3.5$. We then let the dust phase to evolve until $5 \,$Myrs (referred as "\texttt{DustPy}" in Table~\ref{tab:summary models}). For various dust evolution snapshots, we then computed the five dust quantities (Eqs.~\ref{eq:total dust density}, \ref{eq:total dust number density}, \ref{eq:representative grain size}, \ref{eq:representative grain cross-section}, \ref{eq:representative grain mass}), which are used as inputs into our MRI-driven turbulence model. The corresponding $\bar{\alpha}$ can thus be computed for each dust evolution snapshot. Doing so means that we employ our MRI-driven turbulence model following its other mode where the steady-state accretion assumption is relaxed: Computing, on the fly, the effective MRI-induced turbulent parameter, $\bar{\alpha}$, corresponding to any provided gas (e.g., $\Sigma_{\rm{gas}}(r)$), dust (e.g., $\Sigma_{\rm{dust}}(r,a)$) and stellar (e.g., the stellar X-ray luminosity $L_{\rm{XR}}$) properties.

Since gas evolution is turned off, we need to provide the gas surface density, $\Sigma_{\rm{gas}}$, that is used for both the dust evolution and the MRI-driven turbulence model. It is fixed to an input profile, which follows what we refer to as either "Steady-State 2" (SS2) or "LBP" in Table~\ref{tab:summary models} (see the profiles in \textit{Panel b} of Figs.~\ref{fig:Steady_MRI_v_frag1e2_dustpy_optimal_B_field} and \ref{fig:LBP_MRI_v_frag1e2_dustpy_optimal_B_field}, respectively). On the one hand, "SS2" means that $\Sigma_{\rm{gas}}$ follows the gas surface density profile obtained for the steady-state MRI-driven accretion solution corresponding to Model II, with the same MRN-like grain size distribution described in the previous paragraph, and after applying Rayleigh adjustment to it (see Appendix~\ref{appendix:rayleigh adjustment}). On the other hand, "LBP" means that $\Sigma_{\rm{gas}}$ follows the classical radial profile of a power law combined with an exponential cutoff \citep[self-similar solution,][]{1974MNRAS.168..603L} with a total disk gas mass $M_{\rm{disk}} = 0.05 \,$M$_{\star}$, and a critical radius $R_{c} = 80 \,$au. Additionally, we need to provide the $\bar{\alpha}$ used to perform the dust evolution. We fixed it to an input profile (see at $t = 0 \,$yr in \textit{Panel a} of Figs.~\ref{fig:Steady_MRI_v_frag1e2_dustpy_optimal_B_field}, \ref{fig:LBP_MRI_v_frag1e2_dustpy_optimal_B_field} or \ref{fig:LBP_MRI_v_frag1e3_dustpy_optimal_B_field}) obtained by our MRI-driven turbulence model, assuming the same MRN-like grain size distribution, and the same gas surface density profile (following either condition "SS2" or "LBP") as described above.

In the light of the recent laboratory experiments on icy particles \citep[][]{2018MNRAS.479.1273G,2019ApJ...873...58M,2019ApJ...874...60S}, we assume the fragmentation velocity, $v_{\rm{frag}}$, to be radially constant equal to $1 \, $m.s$^{-1}$ in Models IV and V. For completeness, we also adopt the higher value of $10 \, $m.s$^{-1}$ \citep{Wada2011, Gundlach2011, Gundlach2015} in Model VI.

\subsection{Numerical implementation} \label{sect:Numerical implementation}

In all our simulations, the radial grid is computed from $r_{\rm{min}} = 1\,$au to $r_{\rm{max}} = 200\,$au, with $N_{r} = 256$ cells logarithmically spaced. For every radial grid-point $r \in [r_{\rm{min}},r_{\rm{max}}]$, the corresponding vertical grid is computed from the disk mid-plane ($z = 0$) to $z_{\rm{max}}(r) = 5\, H_{\rm{gas}}(r)$, with $N_{z} = 512$ cells linearly spaced. We note that the vertical grid is only used to run our 1+1D MRI-driven turbulence model. 

Regarding the grain size distribution, we always consider a logarithmic grid of grain species whose size range from $a_{\rm{min}}$ (free-parameter depending on the model chosen) to $250 \,$cm (fixed across the models). Furthermore, we consider seven mass bins per mass decade \citep[choice based on the work of][]{1990Icar...83..205O,2014A&A...567A..38D}. For example, the total number of mass bins becomes $N_{m} = 141$ for $a_{\rm{min}} = 0.55 \, \mu$m.

When the dust evolution model is employed, we particularly need to set the dust outer boundary condition. In Model IV, we impose at each time step a constant power-law on the dust surface density, leading to an inflow of dust particles over time consistent with the steady-state MRI-driven accretion assumption for the gas. Indeed, assuming that the gas is in steady-state MRI-driven accretion means that there is a gas and dust reservoir outside the simulation domain due to viscous spreading. In Models V and VI, we impose a floor value on the dust surface density, which essentially prevents the inflow of dust particles from the disk outer regions.

In the dust evolution code \texttt{DustPy}, there are $\delta$ parameters that control the turbulent collision velocities, vertical stirring, and radial diffusion of dust particles. Similar to the documentation, we use the symbols $\delta_{\rm{turb}}$, $\delta_{\rm{vert}}$, and $\delta_{\rm{rad}}$, respectively, for these physical processes. It is up to the user to decide whether such parameters are independent of each other, and depend or not on the $\alpha$-parameter that regulates the gas viscous evolution. In our simulations, we assume the mixing of dust particles to be driven by the MRI and hydrodynamic instabilities captured in our viscous parameter $\bar{\alpha}$. In other words, we assume that $\delta_{\rm{turb}} = \delta_{\rm{vert}} = \delta_{\rm{rad}} = \bar{\alpha}$. Finally, we use \texttt{DustPy} version 0.5.8.

\section{Results} \label{sect:results}

\subsection{The effect of a dust size distribution on the steady-state MRI-driven accretion} \label{sect:the effect of a dust distribution on the steady-state MRI-driven accretion regime}

\begin{figure*}
\centering
\includegraphics[width=\textwidth]{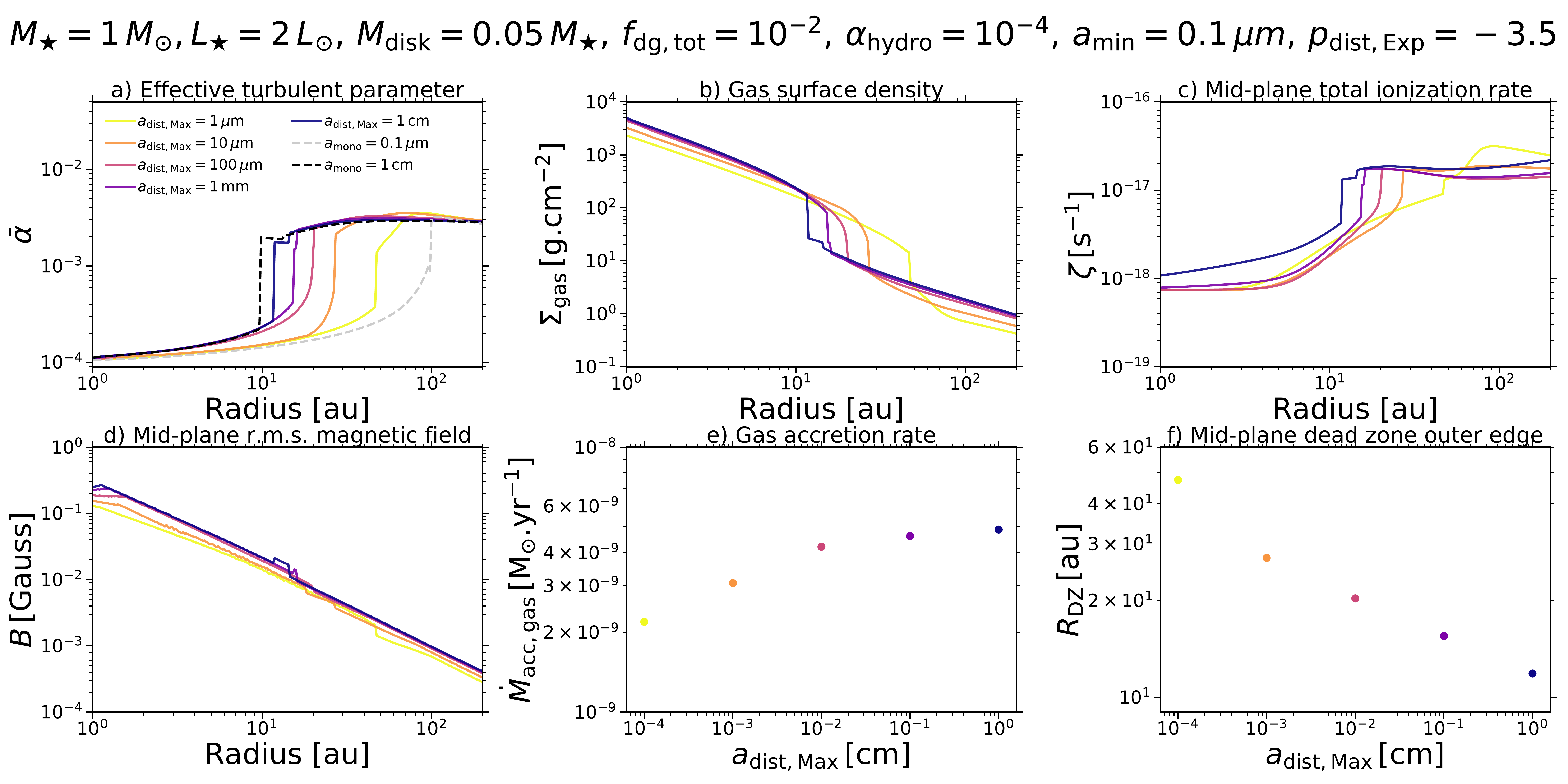}
  \caption{Impact on the steady-state MRI-driven accretion when varying the maximum grain size, $a_{\rm{dist,Max}}$, of the fixed power-law dust size distribution (Model I). Going from solid light-colored to dark-colored lines, $a_{\rm{dist,Max}}$ spans the range from $1 \, \mu$m to $1 \,$cm. The panels show the steady-state radial profiles of several key quantities \citep[see also][for their definition]{Delage2021}, for the model parameters $M_{\star} = 1 \, M_{\odot}$, $L_{\star} = 2 \, L_{\odot}$, $M_{\rm{disk}} = 0.05 \,M_\star$, $f_{\rm{dg,tot}} = 10^{-2}$, $\alpha_{\rm{hydro}} = 10^{-4}$, $a_{\rm{min}} = 0.1 \, \mu$m, and $p_{\rm{dist,Exp}} = -3.5$. \textit{Panel a}: pressure-weighted vertically integrated turbulent parameter, $\bar{\alpha}$. \textit{Panel b}: gas surface density, $\Sigma_{\rm{gas}}$. \textit{Panel c}: mid-plane total ionization rate, $\zeta$. \textit{Panel d}: mid-plane optimal r.m.s. magnetic field strength, $B$. \textit{Panel e}: (constant) gas accretion rate, $\dot{M}_{\rm{acc,gas}}$. \textit{Panel f}: mid-plane radial dead zone outer edge location, $R_{\rm{DZ}}$. For comparison, the dashed gray and black lines in \textit{Panel a} display the steady-state quantity $\bar{\alpha}$ obtained assuming the limiting case of a mono-disperse population of dust with size $a_{\rm{mono}} = 0.1 \, \mu$m and $a_{\rm{mono}} = 1 \,$cm (grain-free case), respectively.}
 \label{fig:Vary_a_distMAX}
\end{figure*}

To study the effect of a dust size distribution on the steady-state MRI-driven accretion investigated in \cite{Delage2021}, we run a set of simulations where we assume the dust size distribution to follow a fixed power-law described by $a_{\rm{min}}$, $a_{\rm{dist,Max}}$, and $p_{\rm{dist,Exp}}$ (see Eq.~\eqref{eq:power-law dust distribution}). In the following, we present the impact of a variation in $a_{\rm{dist,Max}}$, while fixing $a_{\rm{min}} = 0.1 \, \mu$m and $p_{\rm{dist,Exp}} = -3.5$. This set of simulations corresponds to Model I, and the results are presented in Fig.~\ref{fig:Vary_a_distMAX}.

Figs.~\ref{fig:Vary_a_distMAX}a, \ref{fig:Vary_a_distMAX}e and \ref{fig:Vary_a_distMAX}f show that a higher $a_{\rm{dist,Max}}$ leads to stronger MRI-driven turbulence overall, a higher gas accretion rate, and a more compact dead zone. When larger grain sizes are included in the dust distribution, the overall ionization level becomes higher (Fig.~\ref{fig:Vary_a_distMAX}c), leading to enough charged particles in the gas phase for the magnetic field to couple with, mainly due to two reasons: (1) Dust settling becomes more important, which locally leads to an increase in the dust-to-gas mass ratio at the mid-plane, hence promoting the ionization power of radionuclides (dominating the ionization process in the inner regions of the dead zone). (2) The total grain surface area, $A_{\rm{tot}}$, decreases (i.e., grains can less efficiently adsorbe free electrons or ions onto their surfaces), resulting in the gas-phase recombination more easily dominating the recombination process over grain surface adsorption. $A_{\rm{tot}}$ decreases because a fraction of the smaller sizes in the dust distribution is replaced by larger sizes, implying that charged particles in the gas phase encounter, per unit volume and on average, less small dust particles.

Consequently, the MRI can operate with stronger magnetic field strengths on average (Fig.~\ref{fig:Vary_a_distMAX}d). Finding stronger $B$ overall for increasing $a_{\rm{dist,Max}}$ means that ambipolar diffusion becomes less stringent, allowing for MRI-driven turbulence with stronger magnetic field strengths. Furthermore, the MRI can operate closer to the central star, leading the dead zone outer edge to be almost located as twice as close for $a_{\rm{dist,Max}} = 10 \, \mu$m compared to $a_{\rm{dist,Max}} = 1 \, \mu$m. Interestingly, we notice that for the given model parameters the dead zone outer edge is within $10$--$50\,$au, and even $10$--$20\,$au for $a_{\rm{dist,Max}} \geq 100 \, \mu$m. The situation in which the mid-plane would be almost entirely MRI-dead (gray dashed line in Fig.~\ref{fig:Vary_a_distMAX}a), obtained assuming a mono-disperse dust distribution of fixed size $a_{\rm{mono}} = 0.1 \, \mu$m, is thus greatly mitigated when a dust distribution with different sizes is taken into account. We find that the presence of micron-sized particles in the dust size distribution, on top of the submicron-sized particles, prevents the dead zone from extending up to $\sim 100 \,$au for our choice of the magnetic field strength and configuration. 

Figure~\ref{fig:Vary_a_distMAX}a shows that the solid colored lines lie within the dashed gray and black lines, representing the two limiting scenarios for $\bar{\alpha}$ in which all the dust would be either in the form of grains of size $a_{\rm{mono}} = 0.1 \, \mu$m or $a_{\rm{mono}} = 1 \, $cm, respectively. Since the minimum grain size of Model I is $a_{\rm{min}} = 0.1 \, \mu$m and its maximum grain size is $a_{\rm{dist,Max}} = 1 \,$cm, it is expected that the $\bar{\alpha}$ obtained from such dust distributions with different grain sizes can neither represent a less MRI-active scenario than the dashed gray line, nor a more MRI-active scenario than the dashed black line. It is worth mentioning that the models with growth (solid colored lines) converge toward the "grain-free case" for the model parameters considered, corresponding to the dashed black line. The dashed black line mimics the grain-free case because the recombination process occurs in the gas phase rather than onto the grains surface when all grains are of size $1 \,$cm. Indeed, the dashed black line of Fig.~\ref{fig:Vary_a_distMAX}a is identical to the red and purple lines of Fig.~8a of \citet{Delage2021} as well as their blue and yellow lines of Fig.~9a, implying that this is the grain-free steady-state solution because it is independent of the dust properties considered.

From Fig.~\ref{fig:Vary_a_distMAX}, we can infer that the presence of larger sizes in the dust distribution (due to dust growth) substantially impacts the MRI-driven turbulence. Particularly, we expect dust growth to have a major positive impact on the MRI activity in regions where grain surface absorption is the main process for recombination (this recombination regime is highly sensitive to the dust properties), whereas such impact is expected to be weak in regions where the recombination process is dominated by gas-phase recombination (this recombination regime is weakly dependent of the dust properties). Fig.~\ref{fig:Vary_a_distMAX}a even shows that the overall effective turbulence level in the inner regions of the dead zone ($\bar{\alpha}$ for $r \lesssim 10 \,$au) is noticeably different depending on $a_{\rm{dist,Max}}$. This suggests that dust growth is able to change the activity in the MRI-active layer sitting above the dead zone such that the effective turbulence level in the dead zone increases. Furthermore, it seems that dust growth does not need to be very efficient to generate a significant boost in the MRI-driven turbulence. Indeed, the positive feedback obtained from the presence of larger grain sizes on the MRI activity is getting less noticeable once the maximum grain size is larger than $100 \, \mu$m. For example, the gas accretion rate and the dead zone outer edge do not change as much for $a_{\rm{dist,Max}} \geq 100 \, \mu$m as $1 \, \mu\rm{m} \leq a_{\rm{dist,Max}} \leq 100 \, \mu$m (Figs.~\ref{fig:Vary_a_distMAX}e and \ref{fig:Vary_a_distMAX}f).

Here we have only presented the effect of the variation in $a_{\rm{dist,Max}}$ on the steady-state MRI-driven accretion. In Appendix~\ref{appendix:Follow-up on the effect of a dust distribution on the steady-state MRI-driven accretion regime}, we also show the results for the variation in the two remaining parameters $a_{\rm{min}}$ (Model II) and $p_{\rm{dist,Exp}}$ (Model III), respectively. The main conclusion is that increasing any of these three parameters leads to a decrease in the total grain surface area, $A_{\rm{tot}}$; hence stronger MRI-driven turbulence overall, a higher accretion rate, and a more compact dead zone. In other words, we expect that any changes occurring in the dust size distribution due to the evolution of the dust phase should impact the MRI-driven turbulence mainly through the quantity $A_{\rm{tot}}$.

\subsection{The effect of dust evolution on the MRI-driven turbulence} \label{sect:the effect of a dust evolution on the MRI-driven turbulence}

The dust size distribution is expected to deviate from a power-law as the dust phase evolves, even in the gaps and pressure maxima \citep[see e.g.][]{2022arXiv220309266A}. Consequently, we run two dust evolution simulations that we partially couple to our MRI-driven turbulence model, relaxing the assumption of the steady-state accretion, in order to investigate the effect of dust evolution on the MRI activity. These two differ in how the gas surface density profile was chosen (either following the condition"SS2" or "LBP", as explained in Sect.~\ref{sect:Models IV--VI}), as well as in the dust outer boundary condition we adopt.

\subsubsection{"SS2" condition for the gas surface density profile} \label{sect:"SS2" condition for the gas surface density}

\begin{figure*}
\centering
\includegraphics[width=\textwidth]{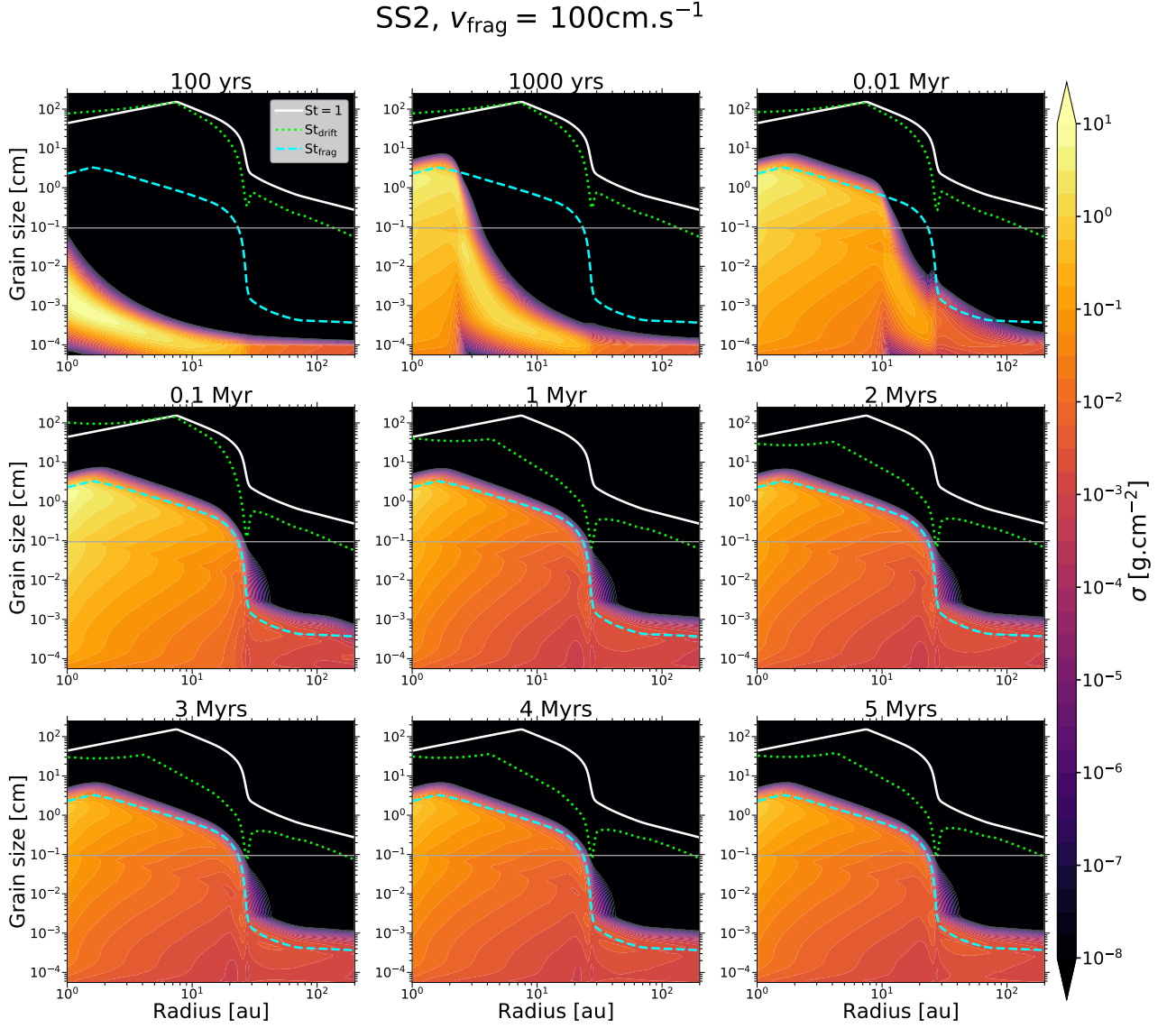}
  \caption{Temporal evolution of the dust surface density distribution per logarithmic bin of grain size $\sigma(r,a)$ (see Eq.~\eqref{eq:sigma}), for Model IV. In each panel, the solid white line shows a Stokes number of unity (radial drift reaches its maximal efficiency), the dotted green line shows the drift limit (Eq.~\eqref{eq:drift_limit}), and the dashed cyan line shows the fragmentation limit (Eq.~\eqref{eq:fragmentation_limit}). The horizontal solid dark gray lines show a grain size of $1 \,$mm.}
 \label{fig:Steady_MRI_v_frag1e2_sigma_dustVSgrainVSradius}
\end{figure*}

\begin{figure*}
\centering
\includegraphics[width=\textwidth]{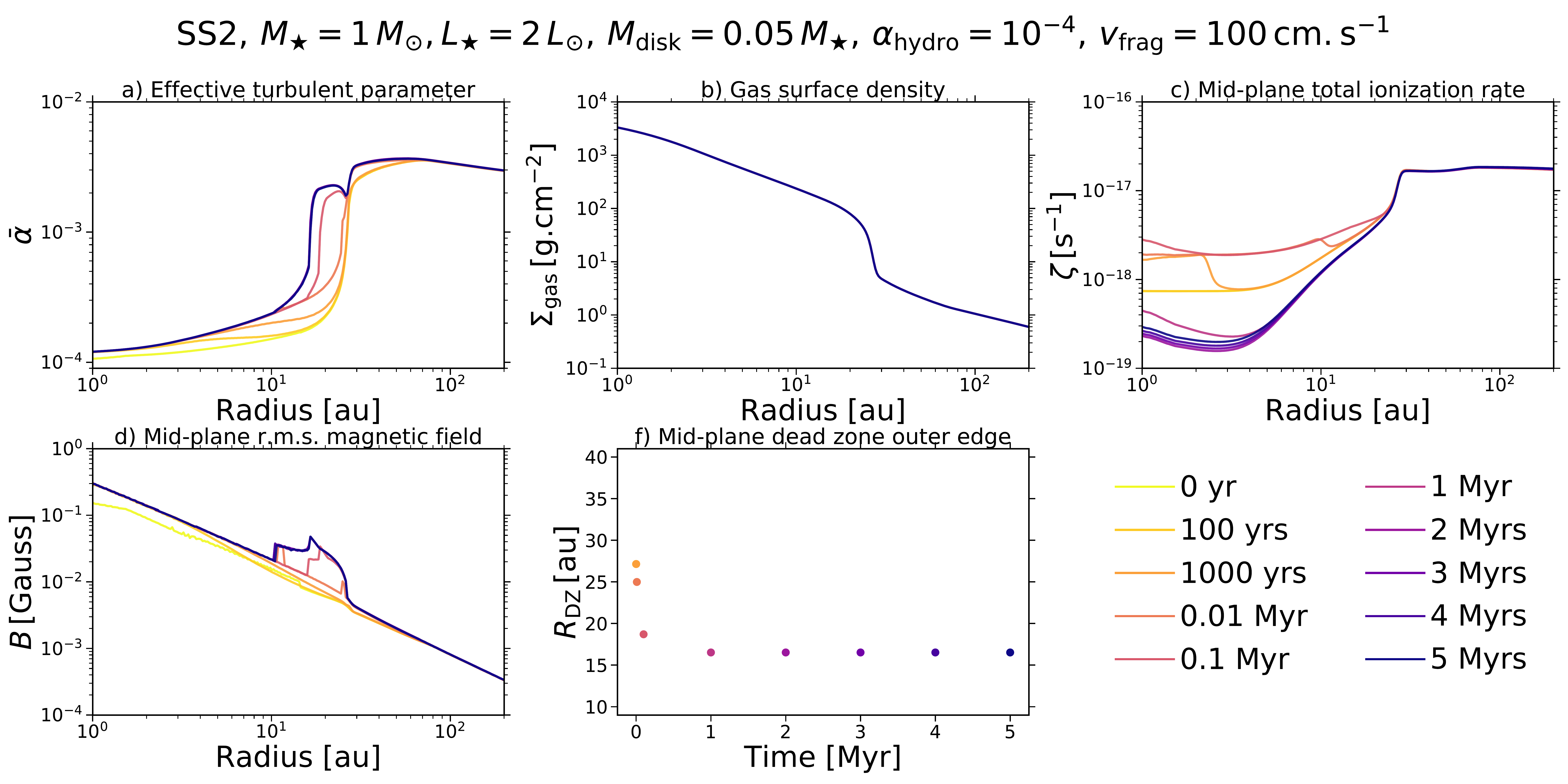}
  \caption{Impact of dust evolution on the MRI-driven turbulence, for Model IV. The panels show the temporal evolution of the same quantities as in Fig.~\ref{fig:Vary_a_distMAX}, except for the gas accretion rate. Also, the gas surface density is now fixed to the displayed input profile in \textit{Panel b}. We emphasize that these quantities do not describe steady-state MRI-driven accretion (unlike Fig.~\ref{fig:Vary_a_distMAX}), since they are re-calculated at each dust evolution snapshot, through partial coupling between the 1D radial dust evolution model employed and our MRI-driven turbulence model. On this note, the corresponding temporal evolution of the five dust quantities used to perform such a coupling are shown in Fig.~\ref{fig:Steady_MRI_v_frag1e2_key_quantities_for_MRI}. Here we note that, in \textit{Panel f}, the dead zone outer edge coincides at $t = 0\,$yr, $t = 100\,$yrs and $t = 1000\,$yrs.
  }
 \label{fig:Steady_MRI_v_frag1e2_dustpy_optimal_B_field}
\end{figure*}

\begin{figure*}
\centering
\includegraphics[width=\textwidth]{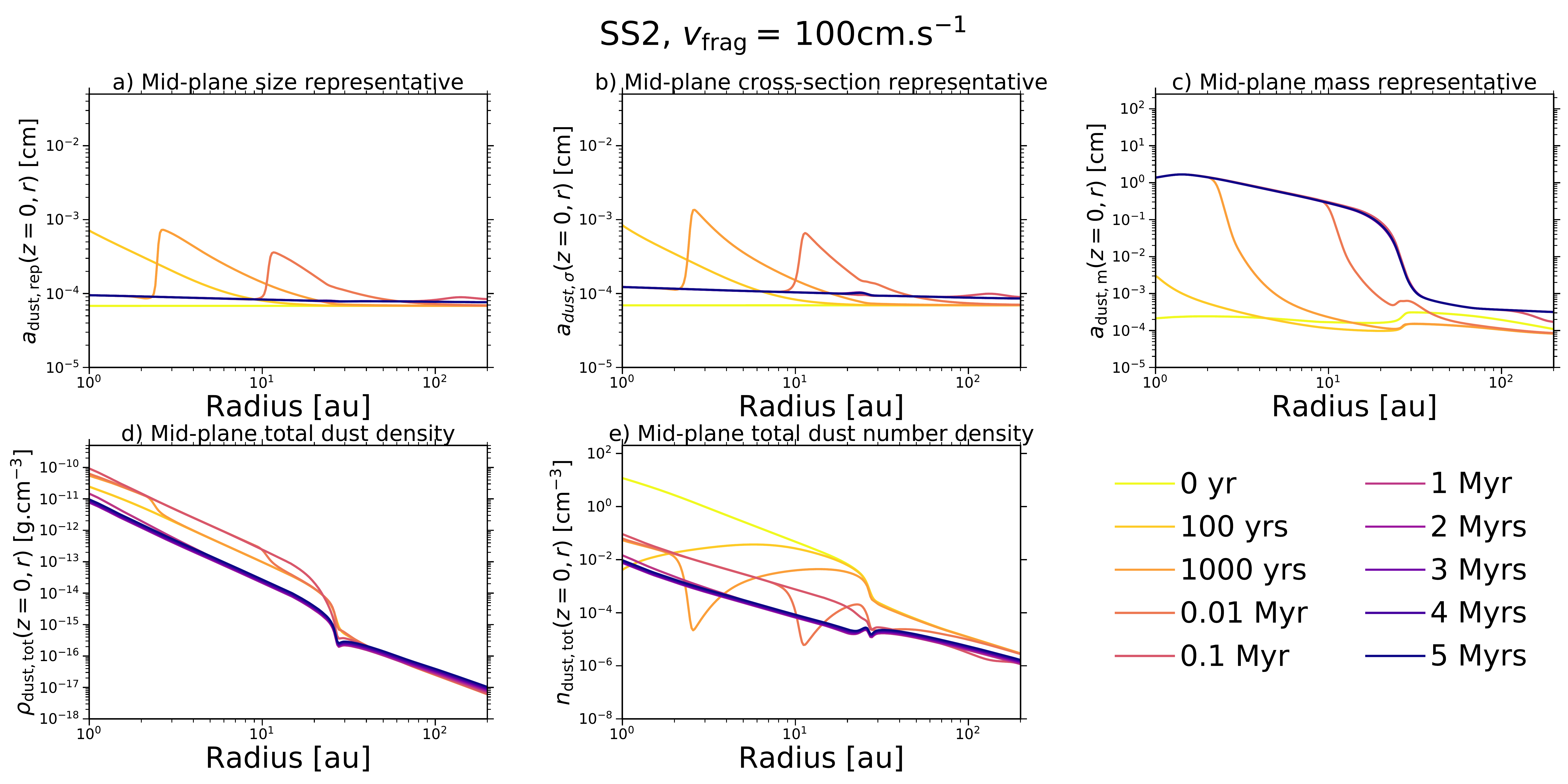}
\caption{Temporal evolution of the five dust quantities used to couple the 1D radial dust evolution model employed with our MRI-driven turbulence model, for Model IV. These quantities are the representative grain size $a_{\rm{dust,rep}}$ (Eq.~\eqref{eq:representative grain size}) in \textit{Panel a}, the representative grain cross-section $\sigma_{\rm{dust,rep}}$ (Eq.~\eqref{eq:representative grain cross-section}), the representative grain mass $m_{\rm{dust,rep}}$ (Eq.~\eqref{eq:representative grain mass}), the total dust density $\rho_{\rm{dust,tot}}$ (Eq.~\eqref{eq:total dust density}) in \textit{Panel d}, and the total dust number density $n_{\rm{dust,tot}}$ (Eq.~\eqref{eq:total dust number density}) in \textit{Panel e}. Instead of displaying $\sigma_{\rm{dust,rep}}$ and $m_{\rm{dust,rep}}$, we show the equivalent grain size $a_{\rm{dust,\sigma}} = \left( \sigma_{\rm{dust,rep}}/\pi \right)^{\frac{1}{2}}$ corresponding to $\sigma_{\rm{dust,rep}}$ (\textit{Panel b}), and the equivalent grain size $a_{\rm{dust,m}} = \left( 3 m_{\rm{dust,rep}}/4 \pi \rho_{\rm{bulk}} \right)^{\frac{1}{3}}$ corresponding to $m_{\rm{dust,rep}}$ (\textit{Panel c}). These equivalent grain sizes better indicate the dominant size when considering the grain cross-section or the grain mass, each of which is important for the MRI calculations (see Sect.~\ref{sect:MRI-driven turbulence}). The panels particularly show the mid-plane radial profiles. We note that $a_{\rm{dust,rep}}$ and $a_{\rm{dust,m}}$ traces the smallest and largest sizes of the dust distribution, respectively, whereas $a_{\rm{dust,\sigma}}$ is the relevant size of the dust distribution involved in the ionization chemistry.}
 \label{fig:Steady_MRI_v_frag1e2_key_quantities_for_MRI}
\end{figure*}

This simulation assumes that the gas surface density profile is stationary, following the steady-state MRI-driven accretion solution corresponding to Model II with a MRN-like dust size distribution ($a_{\rm{min}} = 0.55 \, \mu$m, $a_{\rm{dist,Max}} = 1 \, \mu$m and $p_{\rm{dist,Exp}} = -3.5$; see Appendix~\ref{appendix:variation in the distribution minimum grain size}), and after applying Rayleigh adjustment to it (see Appendix~\ref{appendix:rayleigh adjustment}). The corresponding $\bar{\alpha}$ computed with our MRI-driven turbulence model is used to perform the dust evolution. From the dust phase perspective, therefore, the dead zone is invariant over time, with its outer edge always located at $\sim 27\,$au.

Figure~\ref{fig:Steady_MRI_v_frag1e2_sigma_dustVSgrainVSradius} shows that the dust can grow up to sizes larger than millimeters inside the dead zone ($r \lesssim 27\,$au), whereas the maximum grain size reaches an upper limit of $\sim 10 \,\mu$m outside of it, with an abrupt transition at the dead zone outer edge. This is because the maximum grain size can be limited by either drift when the drift timescale exceeds the growth timescale, or fragmentation when the collision velocity between dust particles exceeds the material fragmentation velocity $v_{\rm{frag}}$ \citep{2008A&A...487L...1B, Birnstiel2009}. For each limit, the time-dependent maximum Stokes number reachable by a dust particle is determined by
\begin{equation} 
    \rm{St}_{\rm{drift}} = \left|\frac{d\ln{P_{\rm{gas,mid}}}}{d\ln{r}}\right|^{-1} \left( \frac{v_K}{c_s} \right)^{2} f_{\rm{dg,tot}},
    \label{eq:drift_limit}
\end{equation}
and
\begin{equation} 
    \rm{St}_{\rm{frag}} = \frac{1}{3 \bar{\alpha}} \left( \frac{v_{\rm{frag}}}{c_s} \right)^{2}.
    \label{eq:fragmentation_limit}
\end{equation}  

The $\bar{\alpha}$ used to perform the dust evolution is shown at $t = 0\,$yr in Fig.~\ref{fig:Steady_MRI_v_frag1e2_dustpy_optimal_B_field}a. It has a mean value of $\sim 1.7 \times 10^{-4}$ in the dead zone, and $\sim 3 \times 10^{-3}$ in the MRI-active region, with a sharp increase near the dead zone outer edge. Additionally, Fig.~\ref{fig:Steady_MRI_v_frag1e2_sigma_dustVSgrainVSradius} indicates that the maximum grain size is set by the fragmentation barrier (dashed cyan line), everywhere in the disk for the entire dust evolution simulation. Since $\rm{St}_{\rm{frag}}$ is inversely proportional to $\bar{\alpha}$, and the turbulent collision velocity between grains is proportional to it with $\Delta v_{\rm{turb}} \approx \sqrt{3 \bar{\alpha} / \rm{St}} c_s$ \citep[for $\rm{St} \ll 1$,][]{Ormel2007}, dust particles can grow into larger sizes in the dead zone compared to the MRI-active region.

The turbulence level used to perform the dust evolution is so low in the regions within $30\,$au that grain coagulation is effective during the first stages of dust evolution (from $0\,$yr to $0.1\,$Myr), even leading to a depletion in the submicron-sized particles in some parts of the dead zone (top row of Fig.~\ref{fig:Steady_MRI_v_frag1e2_sigma_dustVSgrainVSradius}). We can better appreciate this depletion by looking at Figs.~\ref{fig:Steady_MRI_v_frag1e2_key_quantities_for_MRI}a and \ref{fig:Steady_MRI_v_frag1e2_key_quantities_for_MRI}b. These panels show the temporal evolution of the mid-plane representative grain size (tracing the smallest sizes of the dust distribution), and the mid-plane equivalent size of the representative grain cross-section (relevant size of the dust distribution involved in the ionization chemistry), respectively. There is a "growth wave" propagating inside-out from the initial time until $0.01\,$Myr, which is attributed to the location where the dust size distribution becomes skewed toward larger sizes because submicron-sized particles grow much quicker than fragmentation can replenish them, as shown by Fig.~\ref{fig:Steady_MRI_v_frag1e2_key_quantities_for_MRI}c. This panel displays the temporal evolution of the mid-plane equivalent size of the representative grain mass (tracing the largest sizes of the dust distribution) in the protoplanetary disk. Since this quantity increases quickly in the dead zone within $0.01\,$Myr, it implies that larger sizes indeed become present in the dust size distribution. The effective growth of submicron-sized particles into larger sizes can also be seen by comparing Figs.~\ref{fig:Steady_MRI_v_frag1e2_key_quantities_for_MRI}d and \ref{fig:Steady_MRI_v_frag1e2_key_quantities_for_MRI}e. While the mid-plane total number dust density ($n_{\rm{dust,tot}}$) decreases within $0.1\,$Myr, the mid-plane total dust density ($\rho_{\rm{dust,tot}}$) increases. Since the total dust content almost remains constant within this period of time, a decrease in $n_{\rm{dust,tot}}$ is primarily related to a decrease in the relative proportion of small dust particles in the dust distributions that have grown into larger sizes (hence the increase of $\rho_{\rm{dust,tot}}$).

The first direct consequence of the initial effective grain growth in the dead zone, within $0.1\,$Myr of dust evolution, is that more dust particles can settle toward the mid-plane. Settling causes an increase in the local dust-to-gas mass ratio at the mid-plane ($\rho_{\rm{gas}}$ is constant here, while $\rho_{\rm{dust,tot}}$ increases as shown by Fig.~\ref{fig:Steady_MRI_v_frag1e2_key_quantities_for_MRI}d). Consequently, the ionization power of radionuclides increases, leading the mid-plane total ionization rate to increase for $r \lesssim 30\,$au (Fig.~\ref{fig:Steady_MRI_v_frag1e2_dustpy_optimal_B_field}c). The second direct consequence is that ambipolar diffusion becomes weaker (particularly where grain surface adsorption dominates the recombination process), since the total grain surface area, $A_{\rm{tot}}$, decreases when the dust grow (see Sect.~\ref{sect:the effect of a dust distribution on the steady-state MRI-driven accretion regime}) or when the dust distribution is skewed toward larger grain sizes (see Appendix~\ref{appendix:variation in the distribution exponent}). In this regard, Fig.~\ref{fig:Steady_MRI_v_frag1e2_dustpy_optimal_B_field}d shows that the MRI is allowed by ambipolar diffusion to have stronger magnetic field strengths for $r \lesssim 30\,$au, as dust evolves from $0\,$yr to $0.1\,$Myr, particularly near the dead zone outer edge. 

From these two consequences, we can understand the temporal evolution of $\bar{\alpha}$ for $r \lesssim 30\,$au, within the first stages of dust evolution (from $0\,$yr to $0.1\,$Myr): A higher ionization level combined with a less stringent ambipolar diffusion result in the MRI activity being able to operate closer to the central star (Fig.~\ref{fig:Steady_MRI_v_frag1e2_dustpy_optimal_B_field}f shows that the dead zone outer edge deceases within $0.1\,$Myr), with stronger turbulence generated for any regions within $30\,$au (Fig.~\ref{fig:Steady_MRI_v_frag1e2_dustpy_optimal_B_field}a). Interestingly, we notice that $\bar{\alpha}$ first varies in the very inner regions of the dead zone due to an increase in the radionuclide ionization rate from dust settling, followed by a variation in the outer regions of the dead zone due to a decrease in the total grain surface area, $A_{\rm{tot}}$, from dust growth. Looking at Fig.~\ref{fig:Steady_MRI_v_frag1e2_sigma_dustVSgrainVSradius}, we find that dust particles reach their maximum sizes between $0.01\,$Myr and $0.1\,$Myr, which coincides with the timescale over which the MRI activity has substantially changed (see Figs.~\ref{fig:Steady_MRI_v_frag1e2_dustpy_optimal_B_field}a and \ref{fig:Steady_MRI_v_frag1e2_dustpy_optimal_B_field}f). This indicates that the timescale over which dust evolution significantly impacts the MRI-driven turbulence is a timescale of local dust growth \citep[see e.g., Eq.~(30) of][for the analytic formula]{2016SSRv..205...41B}.

On another note, we notice that $\bar{\alpha}$ continues to increase while $R_{\rm{DZ}}$ decreases between $0.1\,$Myr and $1\,$Myr (Figs.~\ref{fig:Steady_MRI_v_frag1e2_dustpy_optimal_B_field}a and \ref{fig:Steady_MRI_v_frag1e2_dustpy_optimal_B_field}f). Since the maximum grain sizes have already been reached within $0.1\,$Myr (Fig.~\ref{fig:Steady_MRI_v_frag1e2_sigma_dustVSgrainVSradius}), the explanation for such behaviors no longer lies in dust growth alone. Fig.~\ref{fig:Steady_MRI_v_frag1e2_key_quantities_for_MRI}d shows that $\rho_{\rm{dust,tot}}$ significantly decreases in the regions within $\sim 30\,$au, between $0.1\,$Myr and $1\,$Myr. We can infer that the dust content decreases in these regions, due to radial drift which is faster for particles that have grown to larger sizes. When the dust is removed from the disk, the dust-to-gas mass ratio decreases, which allows the MRI to operate closer to the central star with stronger activity \citep[see Sect.~6.3 of][]{Delage2021}. Indeed, although a decrease in the dust-to-gas mass ratio implies a lower mid-plane total ionization rate for $r \lesssim 30\,$au due to less radionuclides (Fig.~\ref{fig:Steady_MRI_v_frag1e2_dustpy_optimal_B_field}c), the dust is far less efficient in sweeping up free electrons and ions from the gas phase due to a decrease in the total grain surface area, hence resulting in a net increase in the gas ionization degree.  

Despite some dust growth within the first stages of dust evolution for $r \gtrsim 30\,$au, we notice that it is not enough to make a real impact on the temporal evolution of the MRI-induced effective turbulent parameter, the mid-plane total ionization rate or the optimal r.m.s. magnetic field strength (Figs.~\ref{fig:Steady_MRI_v_frag1e2_dustpy_optimal_B_field}a, \ref{fig:Steady_MRI_v_frag1e2_dustpy_optimal_B_field}c, and \ref{fig:Steady_MRI_v_frag1e2_dustpy_optimal_B_field}d). It is because the dominant recombination process in these regions is gas-phase recombination, which is weakly dependent on the dust properties. Consequently, it is expected not to see much change in terms of MRI activity for $r \gtrsim 30\,$au, within $1\,$Myr.

Finally, we need to focus on the late stages of dust evolution ($t > 1\,$Myr), where all the five dust quantities used to couple \texttt{DustPy} with our MRI-driven accretion model reach a quasi-steady-state (Fig.~\ref{fig:Steady_MRI_v_frag1e2_key_quantities_for_MRI}). In the present model, we assumed that the gas surface density is fixed to a steady-state profile (Fig.~\ref{fig:Steady_MRI_v_frag1e2_dustpy_optimal_B_field}b). While the steady-state profile describes the inner inward accreting region ($r \lesssim R_{\rm{t}}$) of the viscously evolving disk, it cannot capture the outer viscously expanding region ($r \gtrsim R_{\rm{t}}$). Here $R_{\rm{t}}$ corresponds to the transition radius at which the gas motions changes from inward to outward in a viscously evolving disk \citep[e.g.,][]{1998ApJ...495..385H}. We thus expect the presence of an outer region outside our simulation domain ($r \gtrsim 200\,$au), which can feed the disk ($1 \lesssim r \lesssim 200\,$au) of submicron- and micron-sized dust particles. Here we aimed to mimic this situation by choosing the dust outer boundary condition such that there is an inflow of (small) dust particles through the outer boundary of the domain. Consequently, the fact that the dust reaches a quasi-steady-state for $t > 1\,$Myr appears to be directly linked to such a choice. It leads the MRI-driven turbulence and the dead zone outer edge to be roughly constant from $1\,$Myr all the way until $5\,$Myr of dust evolution. Particularly, the apparent temporal dead zone stability whilst the dust evolves is thus a mere artifact of our choice for the dust boundary condition in this case. It is worth noting that the radioactive decay of $^{26}$Al has been ignored in the present paper, which means that the radionuclide ionization rate for $t > 1\,$Myr is expected to be smaller than what we assumed \citep[see Eq.~15 of][]{Delage2021}. However, it is not too bothersome because \cite{Delage2021} (see, e.g. their Fig.~C.1) showed that the ionization from radionuclides only dominates the total ionization rate deep within the dead zone (radially and vertically). Consequently, the location of the dead zone outer edge should not significantly vary, in the cases considered, if we account for the decrease over time in the radionuclide ionization rate due to the decay of $^{26}$Al. This statement also holds for the time evolution for $t > 1\,$Myr seen in the next section.

Although the coupling between dust evolution and our MRI-driven turbulence model is only partial in this paper (see Sect.~\ref{sect:Models IV--VI}), it is clear that dust evolution has a significant impact on the MRI activity: dust settling, grain coagulation, and fragmentation drive the change in the gas ionization degree, hence on $\bar{\alpha}$ at each radius $r$, on a timescale of local dust growth. In the specific case of Model IV, it is thus expected that dust evolution drives the gas away from the assumed steady-state MRI-driven accretion (condition "SS2" for the gas surface density profile), in the regions within $\sim 30\,$au, by substantially changing the effective turbulent parameter $\bar{\alpha}$ within $0.1\,$Myr. Indeed, any changes in $\bar{\alpha}$ due to dust evolution directly modifies the 1D advection-diffusion equation for the gas. We expect this change to carry on for the next steps of the dust evolution calculations because it depends on $\bar{\alpha}$ and the gas properties. Consequently, we find that the steady-state MRI-driven accretion that we impose on the gas is actually not physically consistent for $r \lesssim 30\,$au.

\subsubsection{"LBP" condition for the gas surface density profile}\label{sect:"LBP" condition for the gas surface density}

\begin{figure*}
\centering
\includegraphics[width=\textwidth]{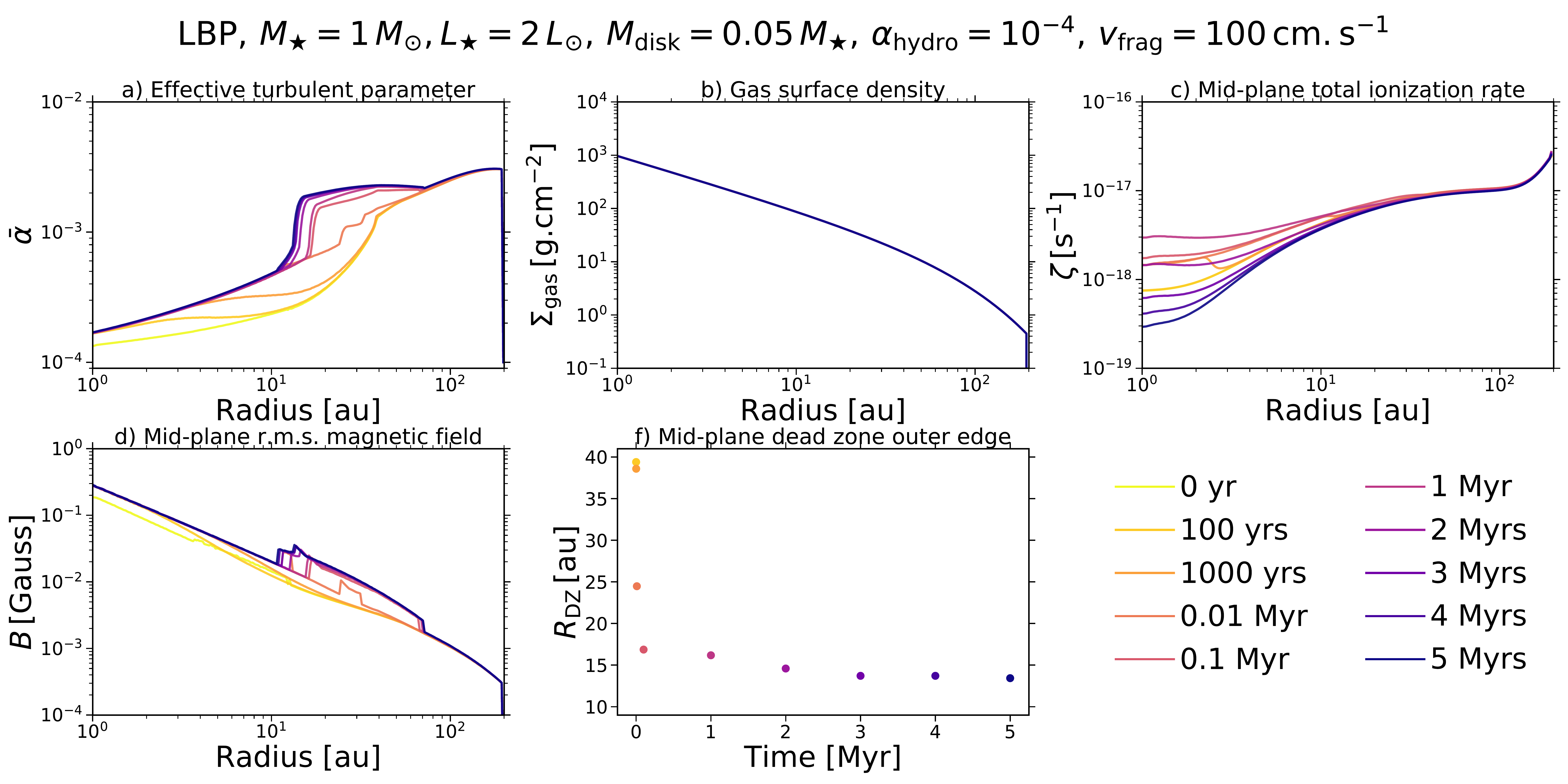}
  \caption{Same as in Fig.~\ref{fig:Steady_MRI_v_frag1e2_dustpy_optimal_B_field}, except for Model V. The corresponding temporal evolution of the five dust quantities is shown in Fig.~\ref{fig:LBP_MRI_v_frag1e2_key_quantities_for_MRI}. The gas surface density (\textit{Panel b}) is fixed to the input profile corresponding to the classical self-similar solution \citep[][]{1974MNRAS.168..603L}, with a total disk gas mass $M_{\rm{disk}} = 0.05 \,$M$_{\star}$, and a critical radius $R_{c} = 80 \,$au. As a result, the gas is no longer in a steady-state MRI-driven accretion at $t = 0 \,$yr, unlike Model IV. Here we note that, in \textit{Panel f}, the dead zone outer edge coincides at $t = 0\,$yr and $t = 100\,$yrs.}
 \label{fig:LBP_MRI_v_frag1e2_dustpy_optimal_B_field}
\end{figure*}

\begin{figure*}
\centering
\includegraphics[width=\textwidth]{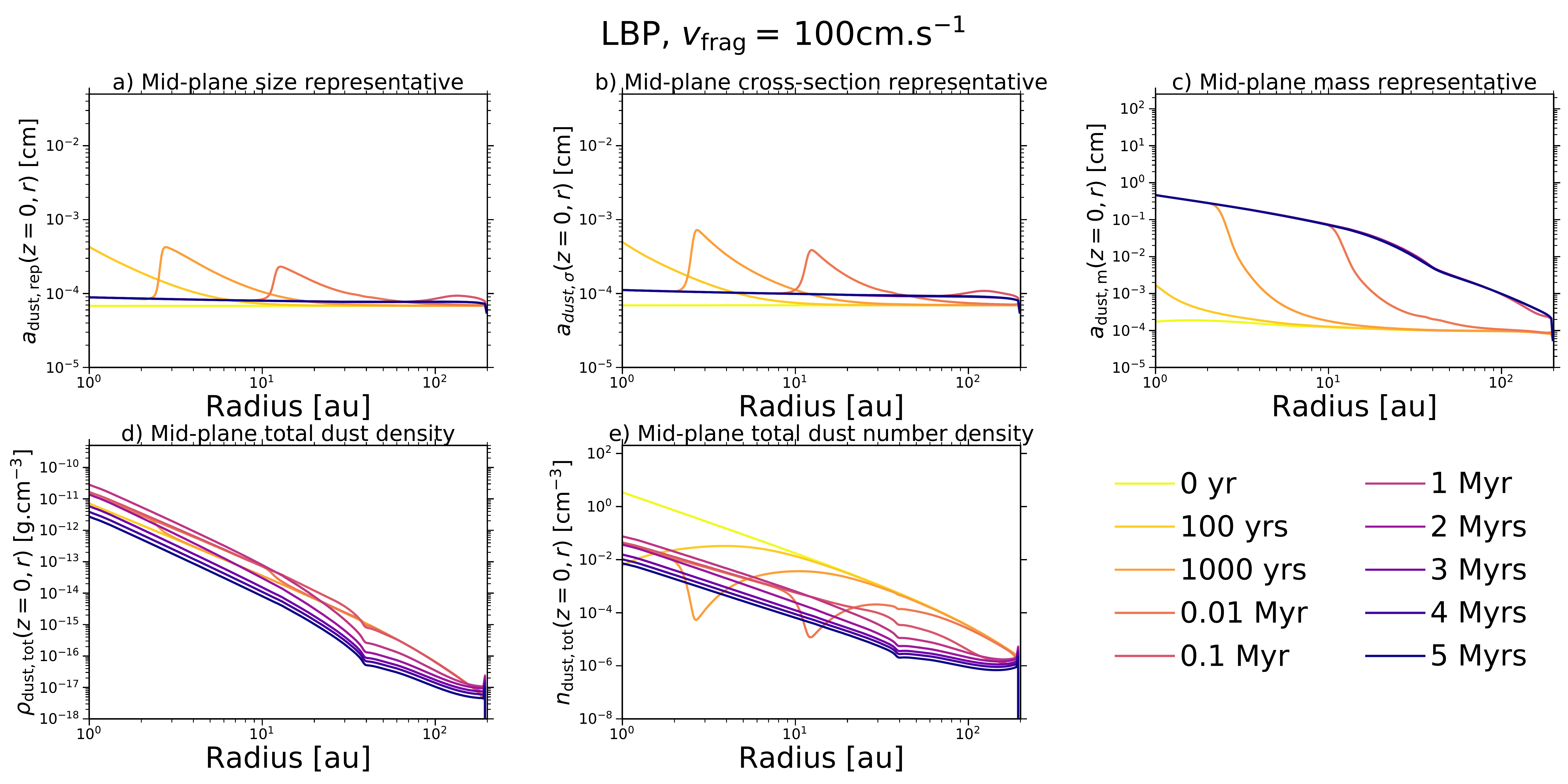}
  \caption{Same as in Fig.~\ref{fig:Steady_MRI_v_frag1e2_key_quantities_for_MRI}, except for Model V.}
 \label{fig:LBP_MRI_v_frag1e2_key_quantities_for_MRI}
\end{figure*}

Here the goal is to see whether the results drawn in the previous section still hold with a different set of assumptions for the gas surface density profile and the dust outer boundary condition. In this case, we assume that the gas surface density profile follows the classical self-similar solution with a total disk gas mass $M_{\rm{disk}} = 0.05 \,$M$_{\star}$, and a critical radius $R_{c} = 80 \,$au (Fig.~\ref{fig:LBP_MRI_v_frag1e2_dustpy_optimal_B_field}b). Here we emphasize again that the gas surface density profile does not evolve with time, so that we can solely focus on the effect of dust evolution on the MRI-driven turbulence. The dust outer boundary condition is chosen such that there is no inflow of small dust particles in the simulation domain, unlike Model IV. Additionally, the $\bar{\alpha}$ used to perform the dust evolution is now derived from the gas surface density profile mentioned above by our MRI-driven turbulence model (see Fig.~\ref{fig:LBP_MRI_v_frag1e2_dustpy_optimal_B_field}a at $t = 0 \,$yr). From the dust perspective, it means that the dead zone is still invariant over time, but with its outer edge now always located at $\sim 40\,$au.

Similar to what we find in the previous section, dust growth is fragmentation-limited everywhere in the disk and throughout its whole evolution, with dust particles growing into larger sizes in the dead zone ($r \lesssim 40\,$au) compared to the MRI-active region (see Fig.~\ref{fig:LBP_MRI_v_frag1e2_sigma_dustVSgrainVSradius}). The main difference, though, is that the particles do not reach sizes as large in the dead zone (there are less grains of size $a \geq 1\,$mm than Model IV), but do reach larger sizes in the MRI-active region (grains can be as large as $\sim 100\,\mu$m), resulting in a much smoother transition in the dust surface density per logarithmic bin of grain size ($\sigma(r,a)$) at the dead zone outer edge. This can be explained by the fact that the $\bar{\alpha}$ used to perform the dust evolution in Model V is on average higher in the dead zone and lower in the MRI-active region than the one used in Model IV, with a much smoother transition at the dead zone outer edge. 

Another common feature between the results of this case (Model V) and Model IV is the effective grain coagulation in the regions within $40\,$au, during the first stages of dust evolution (from $0\,$yr to $0.1\,$Myr). Specifically, it results in a similar growth wave attributed to the depletion of small dust particles that grow into larger sizes (Figs.~\ref{fig:LBP_MRI_v_frag1e2_key_quantities_for_MRI}a, \ref{fig:LBP_MRI_v_frag1e2_key_quantities_for_MRI}b and \ref{fig:LBP_MRI_v_frag1e2_key_quantities_for_MRI}c). In the same fashion as in Sect.~\ref{sect:"SS2" condition for the gas surface density}, effective grain coagulation leads to: (1) more settling of dust particles toward the mid-plane ($\rho_{\rm{dust,tot}}$ increases as shown in Fig.~\ref{fig:LBP_MRI_v_frag1e2_key_quantities_for_MRI}d), allowing for a higher mid-plane ionization rate in the regions within $40\,$au (Fig.~\ref{fig:LBP_MRI_v_frag1e2_dustpy_optimal_B_field}c); (2) less stringent ambipolar diffusion, implying that the MRI can operate with stronger magnetic field strengths ($B$ increases as shown by Fig.~\ref{fig:LBP_MRI_v_frag1e2_dustpy_optimal_B_field}d). Consequently, the MRI activity is able to operate closer to the central star as the dust phase evolves (Fig.~\ref{fig:LBP_MRI_v_frag1e2_dustpy_optimal_B_field}f shows that $R_{\rm{DZ}}$ decreases over time), with stronger turbulence generated in the regions within $40\,$au (Fig.~\ref{fig:LBP_MRI_v_frag1e2_dustpy_optimal_B_field}a). Similar to Model IV, $\bar{\alpha}$ (hence the dead zone outer edge) undergoes significant variation within $0.1\,$Myr, which corresponds to the timescale over which dust particles have reached their maximum size (Fig.~\ref{fig:LBP_MRI_v_frag1e2_sigma_dustVSgrainVSradius}).
This again suggests that the timescale over which dust evolution impacts the MRI-driven turbulence is determined by the timescale of local dust growth. 

Comparing Figs.~\ref{fig:LBP_MRI_v_frag1e2_dustpy_optimal_B_field}a, \ref{fig:LBP_MRI_v_frag1e2_dustpy_optimal_B_field}f with Figs.~\ref{fig:Steady_MRI_v_frag1e2_dustpy_optimal_B_field}a, \ref{fig:Steady_MRI_v_frag1e2_dustpy_optimal_B_field}f, respectively, we notice that the temporal evolution of $\bar{\alpha}$ and $R_{\rm{DZ}}$ between $0.1\,$Myr and $1\,$Myr is less pronounced in Model V compared to Model IV. Indeed, the dust particles within $40\,$au grow into smaller sizes compared to Model IV, which makes their radial drift slower. As a result, they can be present in the disk for a longer period of time, meaning that the removal of the dust content is delayed: Fig.~\ref{fig:LBP_MRI_v_frag1e2_key_quantities_for_MRI}d shows that $\rho_{\rm{dust,tot}}$ still increases between $0.1\,$Myr and $1\,$Myr and only starts decreasing from $1\,$Myr, while it decreases from $0.1\,$Myr in the case of Model IV as shown by Fig.~\ref{fig:Steady_MRI_v_frag1e2_key_quantities_for_MRI}d.

In regions of the disk beyond $\sim 40\,$au (MRI-active region from the dust perspective in Model V), Fig.~\ref{fig:LBP_MRI_v_frag1e2_dustpy_optimal_B_field} shows that, within $1\,$Myr, the effective MRI-induced turbulent parameter, the mid-plane total ionization rate, and the optimal r.m.s. magnetic field strength vary a bit more than in Model IV. We explain this behavior by noticing that the dust particles can grow into larger sizes in the MRI-active region of Model V, especially near the dead zone outer edge ($r \sim 40\,$au). This region marks the transition for the recombination process between grain surface adsorption and gas-phase, and is therefore more sensitive to the dust properties compared to the outer regions. For $r \gtrsim 80\,$au, though, we can see that the MRI activity is pretty much steady within $1\,$Myr. This is expected since $R_{c} = 80\,$au corresponds to the critical radius where the gas surface density profile drops exponentially, and therefore where the gas-phase is the main channel for recombination due to a high gas ionization degree.

To complete the comparison between Model IV and Model V, we now need to discuss the temporal evolution of the MRI-driven turbulence during the late stages of dust evolution ($t > 1\,$Myr). In the previous section, we saw that $\bar{\alpha}$ and $R_{\rm{DZ}}$ were roughly steady after $1\,$Myr of dust evolution, since the five dust quantities used to couple \texttt{DustPy} with our MRI-driven accretion model reach a quasi-steady-state due to the steady influx of dust from the outer boundary. In Model V, though, $\rho_{\rm{dust,tot}}$ and $n_{\rm{dust,tot}}$ keep decreasing from $1\,$Myr all the way until $5\,$Myrs (Figs.~\ref{fig:LBP_MRI_v_frag1e2_key_quantities_for_MRI}d and \ref{fig:LBP_MRI_v_frag1e2_key_quantities_for_MRI}e), since the dust content is gradually removed by radial drift which is no longer compensated for by an inflow of small particles in the simulation domain as in Model IV. This implies that $\bar{\alpha}$ continues to increase while $R_{\rm{DZ}}$ decreases between $1\,$Myr and $3\,$Myr because the MRI can operate closer to the central star with stronger activity for decreasing dust-to-gas mass ratio (as explained in the previous section). Nonetheless, a salient result of Model V is the temporal evolution of the MRI activity between $3\,$Myrs and $5\,$Myrs. Although the dust keeps being removed from the disk, $\bar{\alpha}$ and $R_{\rm{DZ}}$ become stationary (Figs.~\ref{fig:LBP_MRI_v_frag1e2_dustpy_optimal_B_field}a and \ref{fig:LBP_MRI_v_frag1e2_dustpy_optimal_B_field}f). It thus suggests that the MRI activity in the disk becomes weakly dependent on the dust properties when the dust content drops below a certain threshold, which is caused by a lack of dust particles to efficiently sweep up free electrons and ions from the gas phase. In other words, gas-phase recombination dominates in most of the protoplanetary disk after $3\,$Myrs, and the dust has no longer a significant impact on the ionization chemistry. The disk dust-gas mixture thus behaves as a grain-free plasma after $3\,$Myrs, and the dead zone outer edge becomes stationary because the MRI activity evolution becomes primarily controlled by the gas which is not evolving here. This result suggests that the dead zone may potentially be able to survive the protoplanetary disk evolution over a few million years when the MRI is the main driver for the disk accretion. We further discuss this idea in Sect.~\ref{sect:potential long-lived state of the dead zone in protoplanetary disks}.

The results of this section emphasize that dust evolution has a significant impact on the MRI activity, regardless of the assumptions made for the gas surface density profile or the dust outer boundary condition. Particularly, the MRI-induced $\bar{\alpha}$ undergoes substantial change within the timescale over which the dust particles grow.

\section{Discussion} \label{sect:discussion}

\subsection{The effect of the fragmentation velocity} \label{sect:the effect of the fragmentation velocity}

\begin{figure*}
\centering
\includegraphics[width=\textwidth]{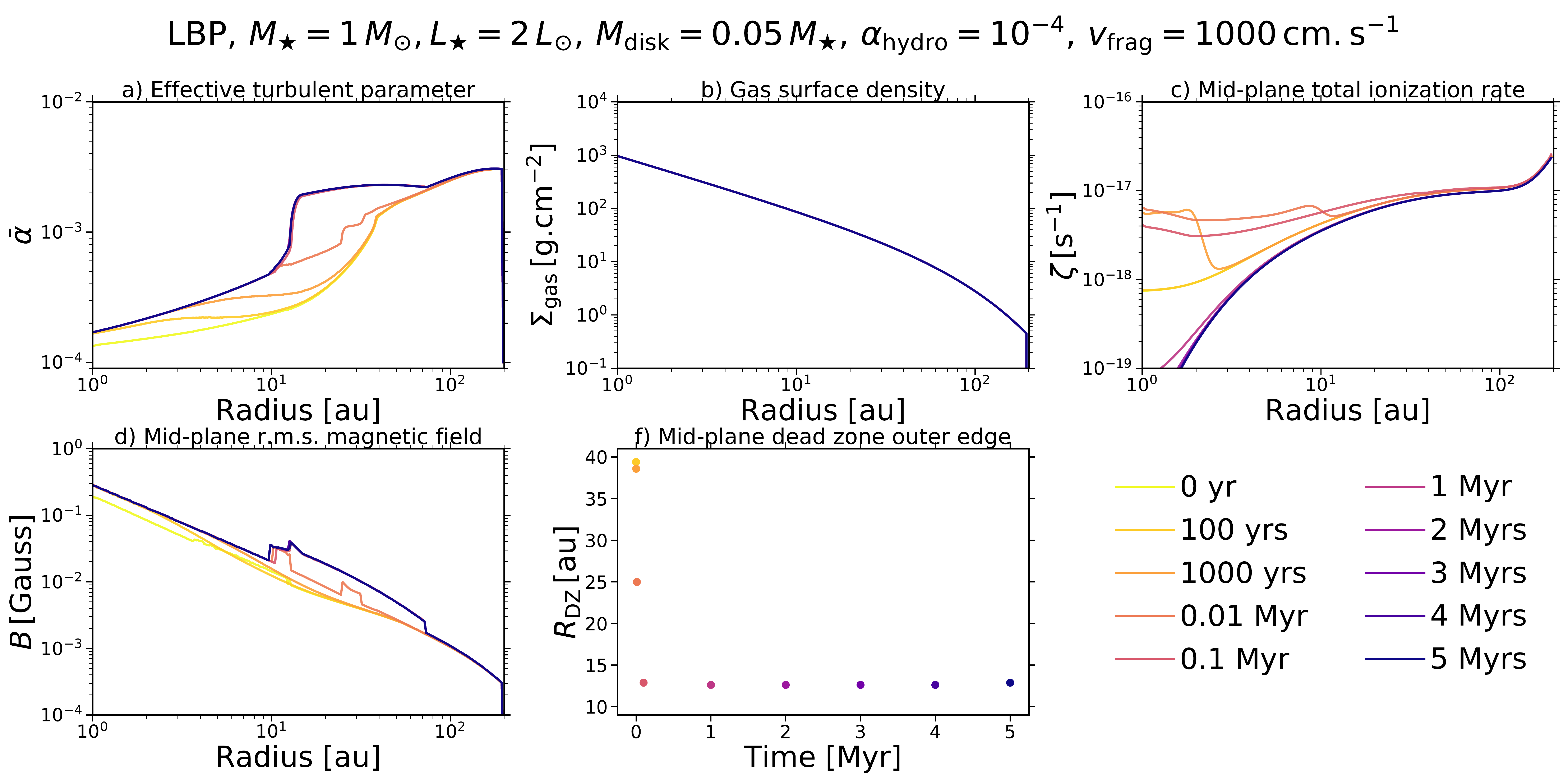}
  \caption{Same as in Fig.~\ref{fig:LBP_MRI_v_frag1e2_dustpy_optimal_B_field}, except for Model VI. The corresponding temporal evolution of the five dust quantities is shown in Fig.~\ref{fig:LBP_MRI_v_frag1e3_key_quantities_for_MRI}. Here we note that, in \textit{Panel f}, the dead zone outer edge coincides at $t = 0\,$yr and $t = 100\,$yrs.}
 \label{fig:LBP_MRI_v_frag1e3_dustpy_optimal_B_field}
\end{figure*}

\begin{figure*}
\centering
\includegraphics[width=\textwidth]{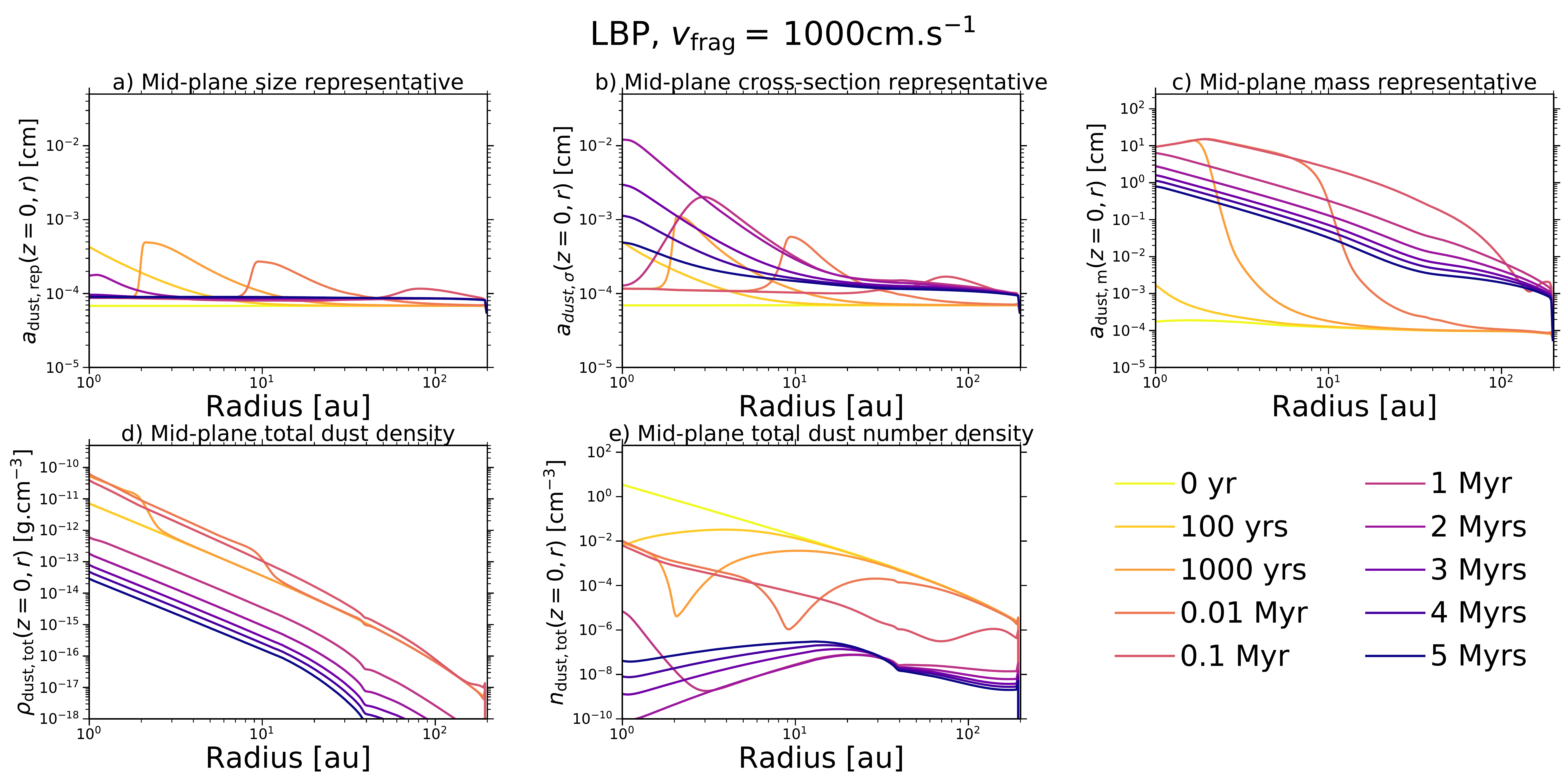}
  \caption{Same as in Fig.~\ref{fig:Steady_MRI_v_frag1e2_key_quantities_for_MRI}, except for Model VI.}
 \label{fig:LBP_MRI_v_frag1e3_key_quantities_for_MRI}
\end{figure*}

Laboratory experiments of particle collisions are crucial in understanding the growth of grains from interstellar medium micron-sized dust to mm/cm-sized pebbles in protoplanetary disks. First, they demonstrate the potential outcome (sticking, fragmentation, bouncing or mass transfer) after grain collisions with a given initial relative velocity \citep[see e.g.][]{2012A&A...540A..73W,2016SSRv..205...41B}. Second, they help to constrain the velocity threshold (fragmentation velocity $v_{\rm{frag}}$) for either effective growth or destructive collisions \citep[e.g.][]{2000Icar..143..138B,2008ARA&A..46...21B,2015ApJ...812...67K,2020MNRAS.498.1801K}. In the classical picture, it has been commonly thought that amorphous water-ice particles are stickier than silicates \citep{Wada2011, Gundlach2011, Gundlach2015}. Theoretical models employing dust evolution thus usually adopt a fragmentation velocity of $10\,$m.s$^{-1}$ for ice particles. However, recent laboratory experiments show that this threshold velocity sensitively depends on the composition and temperature of the colliding particles. They found that water-ice particles may be as fragile as silicates, or even more so, resulting in the fragmentation velocity potentially being as low as $1\,$m.s$^{-1}$ \citep[][]{2018MNRAS.479.1273G,2019ApJ...873...58M,2019ApJ...874...60S}. In Models IV and V, we therefore have experimented with $v_{\rm{frag}} = 1\,$m.s$^{-1}$. For completeness, Model VI investigates how this parameter impacts our results by taking a value of $10\,$m.s$^{-1}$. We note that, except for $v_{\rm{frag}}$, Model VI has the same setup as Model V. 

For increasing value of the fragmentation velocity ($v_{\rm{frag}}$), grain collisions are less destructive which improves the efficiency of coagulation. The fragmentation barrier thus becomes less stringent, allowing grains to grow into larger particles than Model V, everywhere in the disk (Fig.~\ref{fig:LBP_MRI_v_frag1e3_sigma_dustVSgrainVSradius}). Within $0.1\,$Myr of dust evolution, the grains can grow into such large sizes that their Stokes number reach values close to unity. This results in the dust content being quickly removed from the disk between $0.1\,$Myr and $1\,$Myr due to effective radial drift, hence the quick increase in the MRI activity (Fig.~\ref{fig:LBP_MRI_v_frag1e3_dustpy_optimal_B_field}). Indeed, Fig.~\ref{fig:LBP_MRI_v_frag1e3_key_quantities_for_MRI}d shows that $\rho_{\rm{dust,tot}}$ first increases from $0\,$yr to $0.1\,$Myr (caused by dust growth), then plummets by two orders of magnitude between $0.1\,$Myr and $1\,$Myr. As a result, the dust-to-gas mass ratio significantly decreases within $1\,$Myr ($\rho_{\rm{gas}}$ is stationary), implying that the drift barrier gradually becomes more stringent because $\rm{St}_{\rm{drift}}$ decreases when $f_{\rm{dg,tot}}$ decreases (see Eq.~\eqref{eq:drift_limit}). The main consequence is that dust growth is no longer solely fragmentation-limited, unlike Model V. Instead, the temporal evolution of the dust surface density distribution per logarithmic bin of grain size (Fig.~\ref{fig:LBP_MRI_v_frag1e3_sigma_dustVSgrainVSradius}) shows that it transitions from being fragmentation-limited to drift-limited in the entire protoplanetary disk, in a million-year timescale. The larger grains thus radially drift inward before they can collide and replenish the smaller ones, resulting in the depletion of the latter at a wide range of radii within the dead zone as seen by the dust particles ($r \lesssim 40\,$au), from $1\,$Myr. The lack of small dust particles leads the total dust number density to plummet from $0.1\,$Myr to $t > 1\,$Myr by at least two order of magnitudes (Fig.~\ref{fig:LBP_MRI_v_frag1e3_key_quantities_for_MRI}e), and the grain cross-section representative to be skewed toward particles as large as $100\,\mu$m in the disk inner regions (Fig.~\ref{fig:LBP_MRI_v_frag1e3_key_quantities_for_MRI}b). 

Since the rapid drift of the larger particles significantly reduces the dust content in the protoplanetary disk, as well as prevents the smaller ones to be replenished efficiently, the total grain surface area is significantly reduced after $0.1\,$Myr. Consequently, gas-phase recombination becomes the main channel after $0.1\,$Myr, which makes the MRI activity weakly dependent on the dust properties. Particularly, it leads the effective MRI-induced turbulent parameter, the dead zone outer edge, and the optimal r.m.s. magnetic field strength to become stationary (Figs.~\ref{fig:LBP_MRI_v_frag1e3_dustpy_optimal_B_field}a, \ref{fig:LBP_MRI_v_frag1e3_dustpy_optimal_B_field}d, and \ref{fig:LBP_MRI_v_frag1e3_dustpy_optimal_B_field}f) because there is no gas evolution accounted for here. This needs to be put in the context of the stationary temporal evolution of the MRI-driven turbulence between $3\,$Myrs and $5\,$Myrs seen in Sect.~\ref{sect:"LBP" condition for the gas surface density}. In the present model, such a stationary behavior occurs much earlier than in Model V because the dust content is removed on a much shorter timescale due to more effective radial drift. In other words, the disk dust-gas mixture behaves as a grain-free plasma much faster for the present model. Once the treatment of dust and MRI calculations is done simultaneously, we thus expect dust evolution to have a less long-term impact on the MRI activity in the regions of the disk with a higher fragmentation velocity.

\subsection{The potential long-lived state of the dead zone in protoplanetary disks} \label{sect:potential long-lived state of the dead zone in protoplanetary disks}

One of our salient results is that, once the full self-consistent treatment of gas and dust evolution with MRI calculations is done, we expect the temporal evolution of the MRI-driven turbulence to be controlled first by dust evolution, then gas evolution. Indeed, we saw in 
Sects.~\ref{sect:"SS2" condition for the gas surface density} and \ref{sect:"LBP" condition for the gas surface density} that the MRI-induced $\bar{\alpha}$ changes on a timescale of local dust growth, which is significantly shorter than the viscous evolution timescale in general. As long as there is enough dust particles in the disk to dominate the recombination process for the ionization chemistry, we thus expect the MRI activity evolution to be controlled by dust evolution. Once it is no longer the case, the MRI activity evolution is then expected to be controlled by gas evolution and occurs on a viscous evolution timescale, since the dust would no longer have a significant feedback on the ionization chemistry and the disk dust-gas mixture would behave as a grain-free plasma. 

The timescale marking the transition from "dust-dominated" to "gas-dominated" MRI-driven turbulence depends on the physical properties of the protoplanetary disk. For instance, if the disk initially has a low dust content, the transition is expected to happen earlier in the disk lifetime. Conversely, if the disk initially has a high dust content or if it has pressure bumps (non-smooth disk), this transition is expected to occur over a longer timescale. Regarding the latter, it is commonly accepted that pressure bumps are a possible explanation for the observed disk substructures, and can be created by various mechanisms such as embedded massive planets \citep[see e.g.,][]{Zhu_2012, Pinilla_2012, 2015A&A...573A...9P, 2015ApJ...809L...5D, 2021A&A...656A.150P}, dust particle growth by condensation near the ice lines \citep[e.g.,][]{2017A&A...600A.140S}, or magnetic disk winds \citep[e.g.,][]{2017MNRAS.468.3850S,2018MNRAS.477.1239S,2019MNRAS.484..107S,2022MNRAS.tmp.1814H}. If there are pressure bumps in the disk, the main mechanism removing the dust (i.e., radial drift) is not as efficient because a substantial amount of particles can be trapped there. As a result, pressure bumps allow the dust to be in the disk for a longer period of time, hence delaying when the transition from dust-dominated to gas-dominated MRI-driven turbulence occurs.  

In Sects.~\ref{sect:"LBP" condition for the gas surface density} and \ref{sect:the effect of the fragmentation velocity}, we saw that the dead zone shrinks over time due to dust evolution ($R_{\rm{DZ}}$ decreases), and eventually becomes stationary after $3\,$Myrs of disk evolution for Model V, and $0.1\,$Myr for Model VI. Unlike Model IV (Sect.~\ref{sect:"SS2" condition for the gas surface density}), these models have no inflow of small particles feeding the outer boundary of the simulation domain. The stationary nature of the dead zone comes from the fact that the disk dust-gas mixture eventually behaves as a grain-free plasma after some time of evolution, where the MRI activity evolution is primarily controlled by the gas which is not evolving here. Indeed, Figs.~\ref{fig:LBP_MRI_v_frag1e2_dustpy_optimal_B_field}f and \ref{fig:LBP_MRI_v_frag1e3_dustpy_optimal_B_field}f show that the mid-plane dead zone radial extent is larger than $\sim 10\,$au, at any time and for both Models V and VI. Consequently, our results show that dust evolution alone does not lead to a complete reactivation of the dead zone in protoplanetary disks. As long as there is no mechanisms that can efficiently ionize the gas in the inner regions of the disk ($r \leq 10\,$au for a Solar-type star) and that the gas dispersal in those regions occurs in timescales of a few million years, the dead zone may potentially be able to survive the protoplanetary disk evolution over a few millions years when the MRI is the main driver for the disk accretion. This supports that a disk evolution model including X-ray photoevaporative dispersal and a dead zone in the inner regions is a feasible idea in order to successfully explain the main observable properties of transitions disks such as extended gaps and high accretion rates \citep[see][]{2021A&A...655A..18G}.

\subsection{The effect of the dust distribution minimum grain size} \label{sect:Effect of the dust distribution minimum grain size}

In the models where we partially coupled dust evolution with MRI calculations (Models IV, V and VI), we used a dust distribution minimum grain size, $a_{\rm{min}}$, of $0.55\,\mu$m. This quantity is a free-parameter in dust evolution models. From the modeling perspective, the choice of $a_{\rm{min}}$ should not matter because the dust particles quickly forget their initial grain size due to coagulation and fragmentation. As a result, running a pure dust evolution simulation with $a_{\rm{min}} = 0.1 \, \mu$m or $a_{\rm{min}} = 0.55 \, \mu$m should not substantially change the dust density output.

However, this is no longer true if MRI calculations are now combined with the dust evolution model. As we saw in Appendix~\ref{appendix:variation in the distribution minimum grain size}, taking either $a_{\rm{min}} = 0.1 \, \mu$m or $a_{\rm{min}} = 0.55 \, \mu$m for the distribution minimum grain size leads to appreciably different outcomes in terms of the MRI activity (particularly, the dead zone morphology and the location of its outer edge). Indeed, if a larger grain size is used for the distribution minimum size, the representative grain size involved in the ionization chemistry is skewed toward larger sizes, implying that the total grain surface area, $A_{\rm{tot}}$, decreases and the overall MRI activity increases (see Figs.~\ref{fig:Vary_a_min}a and \ref{fig:Vary_a_min}f). In the context of a full coupling between dust evolution and MRI calculations, the MRI-driven turbulence would be sensitive to the dust distribution and its minimum grain size at any point in time (except if it is gas-dominated as discussed in the previous section). The choice for the adopted value of $a_{\rm{min}}$ is thus crucial.

\citet[][]{2022arXiv220408506T} recently investigated the effect of monomer size and composition on scattering polarization of dust particles by using an exact light scattering technique. By comparing their simulations to observations, they estimated the monomer radius of dust particles to be no greater than $0.4\,\mu$m for several protoplanetary disks. They even found that a minimum grain size of $0.1$--$0.2\,\mu$m appears to explain the recent polarimetric observations of the disk around HD 142527. Nevertheless, they have not excluded the possibility that the monomers could actually be much smaller than $0.1\,\mu$m. If that were the case, polycyclic aromatic hydrocarbon (PAH), representing the smallest end of a grain size distribution, would need to be considered in the MRI calculations. Counter-intuitively, though, it has been shown by various works that including PAHs in the dust size distribution reduces ambipolar diffusion, hence enhancing the overall MRI activity \citep[e.g.,][]{2011ApJ...739...51B,2016MNRAS.460.2050Z,2020ApJ...900..180M}. Consequently, if one wants to accurately describe the MRI-driven turbulence in protoplanetary disks, further studies are required with the aim to provide realistic constraints on the minimum grain size of the dust distribution.

\section{Summary and Conclusions} \label{sect:summary and conclusions}

In this pilot study, we provide an important step toward a better understanding of the MRI--dust coevolution in protoplanetary disks, with the aim to present a proof of concept that dust evolution ultimately plays a crucial role in the MRI activity. To this end, we divided our analysis into two parts:
First, we studied how a fixed power-law dust size distribution with varying parameters impacts the MRI activity, especially the steady-state MRI-driven accretion described in \citet{Delage2021}, by employing and improving their 1+1D MRI-driven turbulence model. Second, we relaxed the steady-state accretion assumption in this newly improved turbulence model, and partially coupled it to the dust evolution code \texttt{DustPy}. Doing so allows us to unveil, for the first time, some insights about how the evolution of dust (dynamics and grain growth processes combined) and MRI-driven accretion are intertwined on million-year timescales, from a more sophisticated modeling of the gas ionization degree. Our key results can be summarized as follows:
\begin{enumerate}
    \item Dust coagulation and settling lead to a higher gas ionization degree (the recombination rate onto grains decreases), resulting in stronger MRI-driven turbulence as well as a more compact dead zone. On the other hand, fragmentation has an opposite effect because it replenishes the disk in small dust particles which are very efficient in sweeping up free electrons and ions from the gas phase (the recombination rate onto grains increases). Since the dust content of the protoplanetary disk decreases over million years of evolution due to radial drift, the MRI-driven turbulence overall becomes stronger and the dead zone more compact until the disk dust-gas mixture eventually behaves as a grain-free plasma;

    \item The MRI activity evolution (hence the temporal evolution of the MRI-induced $\alpha$-parameter) is controlled by dust evolution and occurs on a timescale of local dust growth, as long as there is enough dust particles in the disk to dominate the recombination process for the ionization chemistry. Once it is no longer the case, the MRI activity evolution is expected to be controlled by gas evolution and occurs on a viscous evolution timescale;  
    
    \item Dust evolution alone does not lead to a complete reactivation of the dead zone, since the dust eventually has no significant impact on the ionization chemistry when the disk dust-gas mixture behaves as a grain-free plasma. Such result suggests that the dead zone may potentially be able to survive the protoplanetary disk evolution over a few million years when the MRI is the main driver for the disk accretion, as long as there is no mechanisms that can efficiently ionize the gas in the inner regions of the disk where the dead zone sits at and that the gas dispersal in those regions occurs in timescales of a few million years;
    
    \item For typical T-Tauri stars, the dead zone outer edge is expected to be located roughly between $10\,$au and $50\,$au during the protoplanetary disk lifetime for our choice of the magnetic field strength and configuration;    
    
    \item The MRI activity evolution in protoplanetary disks is expected to be crucially sensitive to the choice made for the minimum grain size of the dust distribution, especially in the early stages of the disk lifetime when the dust has a significant feedback on the ionization chemistry. Further studies focusing on constraining such minimum grain size are thus fundamental.
\end{enumerate}

The evolution on million-year timescales of the MRI activity in protoplanetary disks is a complex problem that significantly depends on the dust properties and how it evolves (this study) as well as the gas and stellar properties. A comprehensive approach to investigate the potential dust trapping power of the dead zone outer edge thus requires a time-dependent framework where MRI calculations are self-consistently combined with gas, dust and stellar evolution on million-year timescales. Armed with our new framework combining our MRI-driven turbulence model and dust evolution (\texttt{DustPy}), we aim to achieve such a self-consistent model as our next step.

\begin{acknowledgements}

This work made extensive use of the Astropy \citep{2013A&A...558A..33A}, Matplotlib \citep{Hunter:2007}, Numpy \citep{harris2020array}, Scipy \citep{2020SciPy-NMeth} software packages. T.N.D., P.P. and M.G. acknowledge support provided by the Alexander von Humboldt Foundation in the framework of the Sofja Kovalevskaja Award endowed by the Federal Ministry of Education and Research. C-C.Y. is grateful for the support from NASA via the Astrophysics Theory Program (grant number 80NSSC21K0141), NASA via the Emerging Worlds program (grant number 80NSSC20K0347), and NASA via the Theoretical and Computational Astrophysics Networks program (grant number 80NSSC21K0497). S.O. is supported by JSPS KAKENHI Grant Numbers JP18H05438, JP19K03926, JP20H01948, and 20H00182. M.F. acknowledges funding from the European Research Council (ERC) under the European Union’s Horizon 2020 research and innovation program (grant agreement no. 757957). T.B. and S.M.S. acknowledge funding from the European Research Council (ERC) under the European Union's Horizon 2020 research and innovation programme under grant agreement No 714769 and funding by the Deutsche Forschungsgemeinschaft (DFG, German Research Foundation) under grants 361140270 and 325594231.
This research was also supported by the Munich Institute for Astro-, Particle and BioPhysics (MIAPbP) which is funded by the Deutsche Forschungsgemeinschaft (DFG, German Research Foundation) under Germany´s Excellence Strategy -- EXC-2094 -- 390783311.

\end{acknowledgements}

\bibliographystyle{aa} 
\bibliography{Bibliography.bib}

\begin{thebibliography}{107}
\expandafter\ifx\csname natexlab\endcsname\relax\def\natexlab#1{#1}\fi

\bibitem[{{Andama} {et~al.}(2022){Andama}, {Ndugu}, {Anguma}, \&
  {Jurua}}]{2022arXiv220309266A}
{Andama}, G., {Ndugu}, N., {Anguma}, S. .~K., \& {Jurua}, E. 2022, arXiv
  e-prints, arXiv:2203.09266

\bibitem[{Andrews {et~al.}(2016)Andrews, Wilner, Zhu, Birnstiel, Carpenter,
  Pérez, Bai, Öberg, Hughes, Isella, \& et~al.}]{Andrews2016}
Andrews, S.~M., Wilner, D.~J., Zhu, Z., {et~al.} 2016, The Astrophysical
  Journal, 820, L40

\bibitem[{Armitage(2011)}]{Armitage_2011}
Armitage, P.~J. 2011, Annual Review of Astronomy and Astrophysics, 49,
  195–236

\bibitem[{{Armitage}(2019)}]{2019SAAS...45....1A}
{Armitage}, P.~J. 2019, Saas-Fee Advanced Course, 45, 1

\bibitem[{{Astropy Collaboration} {et~al.}(2013){Astropy Collaboration},
  {Robitaille}, {Tollerud}, {Greenfield}, {Droettboom}, {Bray}, {Aldcroft},
  {Davis}, {Ginsburg}, {Price-Whelan}, {Kerzendorf}, {Conley}, {Crighton},
  {Barbary}, {Muna}, {Ferguson}, {Grollier}, {Parikh}, {Nair}, {Unther},
  {Deil}, {Woillez}, {Conseil}, {Kramer}, {Turner}, {Singer}, {Fox}, {Weaver},
  {Zabalza}, {Edwards}, {Azalee Bostroem}, {Burke}, {Casey}, {Crawford},
  {Dencheva}, {Ely}, {Jenness}, {Labrie}, {Lim}, {Pierfederici}, {Pontzen},
  {Ptak}, {Refsdal}, {Servillat}, \& {Streicher}}]{2013A&A...558A..33A}
{Astropy Collaboration}, {Robitaille}, T.~P., {Tollerud}, E.~J., {et~al.} 2013,
  \aap, 558, A33

\bibitem[{{Bai}(2011{\natexlab{a}})}]{2011ApJ...739...50B}
{Bai}, X.-N. 2011{\natexlab{a}}, \apj, 739, 50

\bibitem[{{Bai}(2011{\natexlab{b}})}]{2011ApJ...739...51B}
{Bai}, X.-N. 2011{\natexlab{b}}, \apj, 739, 51

\bibitem[{Bai(2016)}]{Bai_2016b}
Bai, X.-N. 2016, The Astrophysical Journal, 821, 80

\bibitem[{{Bai} \& {Goodman}(2009)}]{2009ApJ...701..737B}
{Bai}, X.-N. \& {Goodman}, J. 2009, \apj, 701, 737

\bibitem[{{Bai} \& {Stone}(2011)}]{2011ApJ...736..144B}
{Bai}, X.-N. \& {Stone}, J.~M. 2011, \apj, 736, 144

\bibitem[{Bai {et~al.}(2016)Bai, Ye, Goodman, \& Yuan}]{Bai_2016a}
Bai, X.-N., Ye, J., Goodman, J., \& Yuan, F. 2016, The Astrophysical Journal,
  818, 152

\bibitem[{{Balbus} \& {Hawley}(1991)}]{1991ApJ...376..214B}
{Balbus}, S.~A. \& {Hawley}, J.~F. 1991, \apj, 376, 214

\bibitem[{{Balbus} \& {Hawley}(1998)}]{1998RvMP...70....1B}
{Balbus}, S.~A. \& {Hawley}, J.~F. 1998, Reviews of Modern Physics, 70, 1

\bibitem[{{Barraza-Alfaro} {et~al.}(2021){Barraza-Alfaro}, {Flock}, {Marino},
  \& {P{\'e}rez}}]{2021arXiv210601159B}
{Barraza-Alfaro}, M., {Flock}, M., {Marino}, S., \& {P{\'e}rez}, S. 2021, arXiv
  e-prints, arXiv:2106.01159

\bibitem[{{Birnstiel} {et~al.}(2009){Birnstiel}, {Dullemond}, \&
  {Brauer}}]{Birnstiel2009}
{Birnstiel}, T., {Dullemond}, C.~P., \& {Brauer}, F. 2009, \aap, 503, L5

\bibitem[{{Birnstiel} {et~al.}(2010){Birnstiel}, {Dullemond}, \&
  {Brauer}}]{Birnstiel2010}
{Birnstiel}, T., {Dullemond}, C.~P., \& {Brauer}, F. 2010, \aap, 513, A79

\bibitem[{{Birnstiel} {et~al.}(2016){Birnstiel}, {Fang}, \&
  {Johansen}}]{2016SSRv..205...41B}
{Birnstiel}, T., {Fang}, M., \& {Johansen}, A. 2016, \ssr, 205, 41

\bibitem[{Blandford \& Payne(1982)}]{10.1093/mnras/199.4.883}
Blandford, R.~D. \& Payne, D.~G. 1982, Monthly Notices of the Royal
  Astronomical Society, 199, 883

\bibitem[{{Blum} \& {Wurm}(2000)}]{2000Icar..143..138B}
{Blum}, J. \& {Wurm}, G. 2000, \icarus, 143, 138

\bibitem[{{Blum} \& {Wurm}(2008)}]{2008ARA&A..46...21B}
{Blum}, J. \& {Wurm}, G. 2008, \araa, 46, 21

\bibitem[{{Brauer} {et~al.}(2008){Brauer}, {Henning}, \&
  {Dullemond}}]{2008A&A...487L...1B}
{Brauer}, F., {Henning}, T., \& {Dullemond}, C.~P. 2008, \aap, 487, L1

\bibitem[{{Delage} {et~al.}(2022){Delage}, {Okuzumi}, {Flock}, {Pinilla}, \&
  {Dzyurkevich}}]{Delage2021}
{Delage}, T.~N., {Okuzumi}, S., {Flock}, M., {Pinilla}, P., \& {Dzyurkevich},
  N. 2022, \aap, 658, A97

\bibitem[{{Dong} {et~al.}(2015){Dong}, {Zhu}, {Rafikov}, \&
  {Stone}}]{2015ApJ...809L...5D}
{Dong}, R., {Zhu}, Z., {Rafikov}, R.~R., \& {Stone}, J.~M. 2015, \apjl, 809, L5

\bibitem[{{Dr{\k{a}}{\.z}kowska} {et~al.}(2014){Dr{\k{a}}{\.z}kowska},
  {Windmark}, \& {Dullemond}}]{2014A&A...567A..38D}
{Dr{\k{a}}{\.z}kowska}, J., {Windmark}, F., \& {Dullemond}, C.~P. 2014, \aap,
  567, A38

\bibitem[{{Dubrulle} {et~al.}(1995){Dubrulle}, {Morfill}, \&
  {Sterzik}}]{1995Icar..114..237D}
{Dubrulle}, B., {Morfill}, G., \& {Sterzik}, M. 1995, \icarus, 114, 237

\bibitem[{{Dzyurkevich} {et~al.}(2010){Dzyurkevich}, {Flock}, {Turner},
  {Klahr}, \& {Henning}}]{2010A&A...515A..70D}
{Dzyurkevich}, N., {Flock}, M., {Turner}, N.~J., {Klahr}, H., \& {Henning}, T.
  2010, \aap, 515, A70

\bibitem[{{Fleming} \& {Stone}(2003)}]{2003ApJ...585..908F}
{Fleming}, T. \& {Stone}, J.~M. 2003, \apj, 585, 908

\bibitem[{{Fleming} {et~al.}(2000){Fleming}, {Stone}, \&
  {Hawley}}]{2000ApJ...530..464F}
{Fleming}, T.~P., {Stone}, J.~M., \& {Hawley}, J.~F. 2000, \apj, 530, 464

\bibitem[{{Flock} {et~al.}(2011){Flock}, {Dzyurkevich}, {Klahr}, {Turner}, \&
  {Henning}}]{2011ApJ...735..122F}
{Flock}, M., {Dzyurkevich}, N., {Klahr}, H., {Turner}, N.~J., \& {Henning}, T.
  2011, \apj, 735, 122

\bibitem[{{Flock} {et~al.}(2015){Flock}, {Ruge}, {Dzyurkevich}, {Henning},
  {Klahr}, \& {Wolf}}]{2015A&A...574A..68F}
{Flock}, M., {Ruge}, J.~P., {Dzyurkevich}, N., {et~al.} 2015, \aap, 574, A68

\bibitem[{{Flock} {et~al.}(2020){Flock}, {Turner}, {Nelson}, {Lyra}, {Manger},
  \& {Klahr}}]{2020ApJ...897..155F}
{Flock}, M., {Turner}, N.~J., {Nelson}, R.~P., {et~al.} 2020, \apj, 897, 155

\bibitem[{{Gammie}(1996)}]{1996ApJ...457..355G}
{Gammie}, C.~F. 1996, \apj, 457, 355

\bibitem[{{G{\'a}rate} {et~al.}(2021){G{\'a}rate}, {Delage}, {Stadler},
  {Pinilla}, {Birnstiel}, {Stammler}, {Picogna}, {Ercolano}, {Franz}, \&
  {Lenz}}]{2021A&A...655A..18G}
{G{\'a}rate}, M., {Delage}, T.~N., {Stadler}, J., {et~al.} 2021, \aap, 655, A18

\bibitem[{{Gundlach} \& {Blum}(2015)}]{Gundlach2015}
{Gundlach}, B. \& {Blum}, J. 2015, \apj, 798, 34

\bibitem[{{Gundlach} {et~al.}(2011){Gundlach}, {Kilias}, {Beitz}, \&
  {Blum}}]{Gundlach2011}
{Gundlach}, B., {Kilias}, S., {Beitz}, E., \& {Blum}, J. 2011, \icarus, 214,
  717

\bibitem[{{Gundlach} {et~al.}(2018){Gundlach}, {Schmidt}, {Kreuzig},
  {Bischoff}, {Rezaei}, {Kothe}, {Blum}, {Grzesik}, \&
  {Stoll}}]{2018MNRAS.479.1273G}
{Gundlach}, B., {Schmidt}, K.~P., {Kreuzig}, C., {et~al.} 2018, \mnras, 479,
  1273

\bibitem[{{Haisch} {et~al.}(2001){Haisch}, {Lada}, \&
  {Lada}}]{2001ApJ...553L.153H}
{Haisch}, Karl~E., J., {Lada}, E.~A., \& {Lada}, C.~J. 2001, \apjl, 553, L153

\bibitem[{Harris {et~al.}(2020)Harris, Millman, van~der Walt, Gommers,
  Virtanen, Cournapeau, Wieser, Taylor, Berg, Smith, Kern, Picus, Hoyer, van
  Kerkwijk, Brett, Haldane, del R{\'{i}}o, Wiebe, Peterson,
  G{\'{e}}rard-Marchant, Sheppard, Reddy, Weckesser, Abbasi, Gohlke, \&
  Oliphant}]{harris2020array}
Harris, C.~R., Millman, K.~J., van~der Walt, S.~J., {et~al.} 2020, Nature, 585,
  357

\bibitem[{{Hartmann} {et~al.}(1998){Hartmann}, {Calvet}, {Gullbring}, \&
  {D'Alessio}}]{1998ApJ...495..385H}
{Hartmann}, L., {Calvet}, N., {Gullbring}, E., \& {D'Alessio}, P. 1998, \apj,
  495, 385

\bibitem[{Hartmann {et~al.}(2016)Hartmann, Herczeg, \&
  Calvet}]{doi:10.1146/annurev-astro-081915-023347}
Hartmann, L., Herczeg, G., \& Calvet, N. 2016, Annual Review of Astronomy and
  Astrophysics, 54, 135

\bibitem[{{Hawley} {et~al.}(1995){Hawley}, {Gammie}, \&
  {Balbus}}]{1995ApJ...440..742H}
{Hawley}, J.~F., {Gammie}, C.~F., \& {Balbus}, S.~A. 1995, \apj, 440, 742

\bibitem[{{Hu} {et~al.}(2022){Hu}, {Li}, {Zhu}, \&
  {Yang}}]{2022MNRAS.tmp.1814H}
{Hu}, X., {Li}, Z.-Y., {Zhu}, Z., \& {Yang}, C.-C. 2022, \mnras
  [\eprint[arXiv]{2203.05629}]

\bibitem[{{Huang} {et~al.}(2018){Huang}, {Andrews}, {Cleeves}, {{\"O}berg},
  {Wilner}, {Bai}, {Birnstiel}, {Carpenter}, {Hughes}, {Isella}, {P{\'e}rez},
  {Ricci}, \& {Zhu}}]{Huang2018}
{Huang}, J., {Andrews}, S.~M., {Cleeves}, L.~I., {et~al.} 2018, \apj, 852, 122

\bibitem[{Hunter(2007)}]{Hunter:2007}
Hunter, J.~D. 2007, Computing in Science \& Engineering, 9, 90

\bibitem[{{Ilgner} \& {Nelson}(2006)}]{2006A&A...445..223I}
{Ilgner}, M. \& {Nelson}, R.~P. 2006, \aap, 445, 223

\bibitem[{{Inutsuka} \& {Sano}(2005)}]{2005ApJ...628L.155I}
{Inutsuka}, S.-i. \& {Sano}, T. 2005, \apjl, 628, L155

\bibitem[{{Kimura} {et~al.}(2020){Kimura}, {Wada}, {Kobayashi}, {Senshu},
  {Hirai}, {Yoshida}, {Kobayashi}, {Hong}, {Arai}, {Ishibashi}, \&
  {Yamada}}]{2020MNRAS.498.1801K}
{Kimura}, H., {Wada}, K., {Kobayashi}, H., {et~al.} 2020, \mnras, 498, 1801

\bibitem[{{Kimura} {et~al.}(2015){Kimura}, {Wada}, {Senshu}, \&
  {Kobayashi}}]{2015ApJ...812...67K}
{Kimura}, H., {Wada}, K., {Senshu}, H., \& {Kobayashi}, H. 2015, \apj, 812, 67

\bibitem[{Klahr \& Bodenheimer(2003)}]{Klahr_2003}
Klahr, H.~H. \& Bodenheimer, P. 2003, The Astrophysical Journal, 582, 869–892

\bibitem[{{Lesur} {et~al.}(2022){Lesur}, {Ercolano}, {Flock}, {Lin}, {Yang},
  {Barranco}, {Benitez-Llambay}, {Goodman}, {Johansen}, {Klahr}, {Laibe},
  {Lyra}, {Marcus}, {Nelson}, {Squire}, {Simon}, {Turner}, {Umurhan}, \&
  {Youdin}}]{2022arXiv220309821L}
{Lesur}, G., {Ercolano}, B., {Flock}, M., {et~al.} 2022, arXiv e-prints,
  arXiv:2203.09821

\bibitem[{Lin \& Pringle(1987)}]{10.1093/mnras/225.3.607}
Lin, D. N.~C. \& Pringle, J.~E. 1987, Monthly Notices of the Royal Astronomical
  Society, 225, 607

\bibitem[{Lin \& Youdin(2015)}]{Lin_2015}
Lin, M.-K. \& Youdin, A.~N. 2015, The Astrophysical Journal, 811, 17

\bibitem[{Lodato \& Rice(2004)}]{Lodato_2004}
Lodato, G. \& Rice, W. K.~M. 2004, Monthly Notices of the Royal Astronomical
  Society, 351, 630–642

\bibitem[{{Lynden-Bell} \& {Pringle}(1974)}]{1974MNRAS.168..603L}
{Lynden-Bell}, D. \& {Pringle}, J.~E. 1974, \mnras, 168, 603

\bibitem[{Manger {et~al.}(2020)Manger, Klahr, Kley, \& Flock}]{Manger_2020}
Manger, N., Klahr, H., Kley, W., \& Flock, M. 2020, Monthly Notices of the
  Royal Astronomical Society, 499, 1841–1853

\bibitem[{{Marchand} {et~al.}(2020){Marchand}, {Tomida}, {Tanaka},
  {Commer{\c{c}}on}, \& {Chabrier}}]{2020ApJ...900..180M}
{Marchand}, P., {Tomida}, K., {Tanaka}, K. E.~I., {Commer{\c{c}}on}, B., \&
  {Chabrier}, G. 2020, \apj, 900, 180

\bibitem[{{Mathis} {et~al.}(1977){Mathis}, {Rumpl}, \&
  {Nordsieck}}]{1977ApJ...217..425M}
{Mathis}, J.~S., {Rumpl}, W., \& {Nordsieck}, K.~H. 1977, \apj, 217, 425

\bibitem[{{Musiolik} \& {Wurm}(2019)}]{2019ApJ...873...58M}
{Musiolik}, G. \& {Wurm}, G. 2019, \apj, 873, 58

\bibitem[{{Nakagawa} {et~al.}(1986){Nakagawa}, {Sekiya}, \&
  {Hayashi}}]{Nakagawa1986}
{Nakagawa}, Y., {Sekiya}, M., \& {Hayashi}, C. 1986, \icarus, 67, 375

\bibitem[{Nelson {et~al.}(2013)Nelson, Gressel, \& Umurhan}]{Nelson_2013}
Nelson, R.~P., Gressel, O., \& Umurhan, O.~M. 2013, Monthly Notices of the
  Royal Astronomical Society, 435, 2610–2632

\bibitem[{{Ohtsuki} {et~al.}(1990){Ohtsuki}, {Nakagawa}, \&
  {Nakazawa}}]{1990Icar...83..205O}
{Ohtsuki}, K., {Nakagawa}, Y., \& {Nakazawa}, K. 1990, \icarus, 83, 205

\bibitem[{{Okuzumi} \& {Hirose}(2011)}]{2011ApJ...742...65O}
{Okuzumi}, S. \& {Hirose}, S. 2011, \apj, 742, 65

\bibitem[{{Okuzumi} \& {Hirose}(2012)}]{2012ApJ...753L...8O}
{Okuzumi}, S. \& {Hirose}, S. 2012, \apjl, 753, L8

\bibitem[{{Okuzumi} {et~al.}(2011{\natexlab{a}}){Okuzumi}, {Tanaka},
  {Takeuchi}, \& {Sakagami}}]{2011ApJ...731...95O}
{Okuzumi}, S., {Tanaka}, H., {Takeuchi}, T., \& {Sakagami}, M.-a.
  2011{\natexlab{a}}, \apj, 731, 95

\bibitem[{{Okuzumi} {et~al.}(2011{\natexlab{b}}){Okuzumi}, {Tanaka},
  {Takeuchi}, \& {Sakagami}}]{2011ApJ...731...96O}
{Okuzumi}, S., {Tanaka}, H., {Takeuchi}, T., \& {Sakagami}, M.-a.
  2011{\natexlab{b}}, \apj, 731, 96

\bibitem[{{Ormel} \& {Cuzzi}(2007)}]{Ormel2007}
{Ormel}, C.~W. \& {Cuzzi}, J.~N. 2007, \aap, 466, 413

\bibitem[{{Ormel} \& {Okuzumi}(2013)}]{2013ApJ...771...44O}
{Ormel}, C.~W. \& {Okuzumi}, S. 2013, \apj, 771, 44

\bibitem[{{Perez-Becker} \& {Chiang}(2011)}]{2011ApJ...735....8P}
{Perez-Becker}, D. \& {Chiang}, E. 2011, \apj, 735, 8

\bibitem[{Pinilla {et~al.}(2012)Pinilla, Benisty, \& Birnstiel}]{Pinilla_2012}
Pinilla, P., Benisty, M., \& Birnstiel, T. 2012, Astronomy and Astrophysics,
  545, A81

\bibitem[{{Pinilla} {et~al.}(2015){Pinilla}, {de Juan Ovelar}, {Ataiee},
  {Benisty}, {Birnstiel}, {van Dishoeck}, \& {Min}}]{2015A&A...573A...9P}
{Pinilla}, P., {de Juan Ovelar}, M., {Ataiee}, S., {et~al.} 2015, \aap, 573, A9

\bibitem[{{Pinilla} {et~al.}(2016){Pinilla}, {Flock}, {Ovelar}, \&
  {Birnstiel}}]{2016A&A...596A..81P}
{Pinilla}, P., {Flock}, M., {Ovelar}, M. d.~J., \& {Birnstiel}, T. 2016, \aap,
  596, A81

\bibitem[{{Pollack} {et~al.}(1994){Pollack}, {Hollenbach}, {Beckwith},
  {Simonelli}, {Roush}, \& {Fong}}]{1994ApJ...421..615P}
{Pollack}, J.~B., {Hollenbach}, D., {Beckwith}, S., {et~al.} 1994, \apj, 421,
  615

\bibitem[{{Pyerin} {et~al.}(2021){Pyerin}, {Delage}, {Kurtovic}, {G{\'a}rate},
  {Henning}, \& {Pinilla}}]{2021A&A...656A.150P}
{Pyerin}, M.~A., {Delage}, T.~N., {Kurtovic}, N.~T., {et~al.} 2021, \aap, 656,
  A150

\bibitem[{Raettig {et~al.}(2013)Raettig, Lyra, \& Klahr}]{Raettig_2013}
Raettig, N., Lyra, W., \& Klahr, H. 2013, The Astrophysical Journal, 765, 115

\bibitem[{{Reg{\'a}ly} {et~al.}(2012){Reg{\'a}ly}, {Juh{\'a}sz}, {S{\'a}ndor},
  \& {Dullemond}}]{2012MNRAS.419.1701R}
{Reg{\'a}ly}, Z., {Juh{\'a}sz}, A., {S{\'a}ndor}, Z., \& {Dullemond}, C.~P.
  2012, \mnras, 419, 1701

\bibitem[{{Rodenkirch} \& {Dullemond}(2022)}]{2022A&A...659A..42R}
{Rodenkirch}, P.~J. \& {Dullemond}, C.~P. 2022, \aap, 659, A42

\bibitem[{{Sano} \& {Stone}(2002{\natexlab{a}})}]{2002ApJ...570..314S}
{Sano}, T. \& {Stone}, J.~M. 2002{\natexlab{a}}, \apj, 570, 314

\bibitem[{{Sano} \& {Stone}(2002{\natexlab{b}})}]{2002ApJ...577..534S}
{Sano}, T. \& {Stone}, J.~M. 2002{\natexlab{b}}, \apj, 577, 534

\bibitem[{{Shakura} \& {Sunyaev}(1973)}]{1973A&A....24..337S}
{Shakura}, N.~I. \& {Sunyaev}, R.~A. 1973, \aap, 500, 33

\bibitem[{{Smoluchowski}(1916)}]{1916ZPhy...17..557S}
{Smoluchowski}, M.~V. 1916, Zeitschrift fur Physik, 17, 557

\bibitem[{{Stammler} \& {Birnstiel}(2022)}]{2022arXiv220700322S}
{Stammler}, S.~M. \& {Birnstiel}, T. 2022, arXiv e-prints, arXiv:2207.00322

\bibitem[{{Stammler} {et~al.}(2017){Stammler}, {Birnstiel}, {Pani{\'c}},
  {Dullemond}, \& {Dominik}}]{2017A&A...600A.140S}
{Stammler}, S.~M., {Birnstiel}, T., {Pani{\'c}}, O., {Dullemond}, C.~P., \&
  {Dominik}, C. 2017, \aap, 600, A140

\bibitem[{{Steinpilz} {et~al.}(2019){Steinpilz}, {Teiser}, \&
  {Wurm}}]{2019ApJ...874...60S}
{Steinpilz}, T., {Teiser}, J., \& {Wurm}, G. 2019, \apj, 874, 60

\bibitem[{{Suriano} {et~al.}(2017){Suriano}, {Li}, {Krasnopolsky}, \&
  {Shang}}]{2017MNRAS.468.3850S}
{Suriano}, S.~S., {Li}, Z.-Y., {Krasnopolsky}, R., \& {Shang}, H. 2017, \mnras,
  468, 3850

\bibitem[{{Suriano} {et~al.}(2018){Suriano}, {Li}, {Krasnopolsky}, \&
  {Shang}}]{2018MNRAS.477.1239S}
{Suriano}, S.~S., {Li}, Z.-Y., {Krasnopolsky}, R., \& {Shang}, H. 2018, \mnras,
  477, 1239

\bibitem[{{Suriano} {et~al.}(2019){Suriano}, {Li}, {Krasnopolsky}, {Suzuki}, \&
  {Shang}}]{2019MNRAS.484..107S}
{Suriano}, S.~S., {Li}, Z.-Y., {Krasnopolsky}, R., {Suzuki}, T.~K., \& {Shang},
  H. 2019, \mnras, 484, 107

\bibitem[{Suzuki \& Inutsuka(2009)}]{Suzuki_2009}
Suzuki, T.~K. \& Inutsuka, S.-i. 2009, The Astrophysical Journal, 691,
  L49–L54

\bibitem[{{Takeuchi} \& {Lin}(2002)}]{Takeuchi2002}
{Takeuchi}, T. \& {Lin}, D.~N.~C. 2002, \apj, 581, 1344

\bibitem[{{Tazaki} \& {Dominik}(2022)}]{2022arXiv220408506T}
{Tazaki}, R. \& {Dominik}, C. 2022, arXiv e-prints, arXiv:2204.08506

\bibitem[{{Turner} {et~al.}(2010){Turner}, {Carballido}, \&
  {Sano}}]{2010ApJ...708..188T}
{Turner}, N.~J., {Carballido}, A., \& {Sano}, T. 2010, \apj, 708, 188

\bibitem[{{Turner} \& {Sano}(2008)}]{2008ApJ...679L.131T}
{Turner}, N.~J. \& {Sano}, T. 2008, \apjl, 679, L131

\bibitem[{{Turner} {et~al.}(2007){Turner}, {Sano}, \&
  {Dziourkevitch}}]{2007ApJ...659..729T}
{Turner}, N.~J., {Sano}, T., \& {Dziourkevitch}, N. 2007, \apj, 659, 729

\bibitem[{Urpin \& Brandenburg(1998)}]{10.1111/j.1365-8711.1998.01118.x}
Urpin, V. \& Brandenburg, A. 1998, Monthly Notices of the Royal Astronomical
  Society, 294, 399

\bibitem[{{van Boekel} {et~al.}(2017){van Boekel}, {Henning}, {Menu}, {de
  Boer}, {Langlois}, {M{\"u}ller}, {Avenhaus}, {Boccaletti}, {Schmid},
  {Thalmann}, {Benisty}, {Dominik}, {Ginski}, {Girard}, {Gisler}, {Lobo Gomes},
  {Menard}, {Min}, {Pavlov}, {Pohl}, {Quanz}, {Rabou}, {Roelfsema}, {Sauvage},
  {Teague}, {Wildi}, \& {Zurlo}}]{vanBoekel2017}
{van Boekel}, R., {Henning}, T., {Menu}, J., {et~al.} 2017, \apj, 837, 132

\bibitem[{{van der Marel} \& {Mulders}(2021)}]{2021arXiv210406838V}
{van der Marel}, N. \& {Mulders}, G. 2021, arXiv e-prints, arXiv:2104.06838

\bibitem[{{Villenave} {et~al.}(2019){Villenave}, {Benisty}, {Dent},
  {M{\'e}nard}, {Garufi}, {Ginski}, {Pinilla}, {Pinte}, {Williams}, {de Boer},
  {Morino}, {Fukagawa}, {Dominik}, {Flock}, {Henning}, {Juh{\'a}sz}, {Keppler},
  {Muro-Arena}, {Olofsson}, {P{\'e}rez}, {van der Plas}, {Zurlo}, {Carle},
  {Feautrier}, {Pavlov}, {Pragt}, {Ramos}, {Sauvage}, {Stadler}, \&
  {Weber}}]{2019A&A...624A...7V}
{Villenave}, M., {Benisty}, M., {Dent}, W.~R.~F., {et~al.} 2019, \aap, 624, A7

\bibitem[{{Villenave} {et~al.}(2020){Villenave}, {M{\'e}nard}, {Dent},
  {Duch{\^e}ne}, {Stapelfeldt}, {Benisty}, {Boehler}, {van der Plas}, {Pinte},
  {Telkamp}, {Wolff}, {Flores}, {Lesur}, {Louvet}, {Riols}, {Dougados},
  {Williams}, \& {Padgett}}]{2020A&A...642A.164V}
{Villenave}, M., {M{\'e}nard}, F., {Dent}, W.~R.~F., {et~al.} 2020, \aap, 642,
  A164

\bibitem[{Virtanen {et~al.}(2020)Virtanen, Gommers, Oliphant, Haberland, Reddy,
  Cournapeau, Burovski, Peterson, Weckesser, Bright, {van der Walt}, Brett,
  Wilson, Millman, Mayorov, Nelson, Jones, Kern, Larson, Carey, Polat, Feng,
  Moore, {VanderPlas}, Laxalde, Perktold, Cimrman, Henriksen, Quintero, Harris,
  Archibald, Ribeiro, Pedregosa, {van Mulbregt}, \& {SciPy 1.0
  Contributors}}]{2020SciPy-NMeth}
Virtanen, P., Gommers, R., Oliphant, T.~E., {et~al.} 2020, Nature Methods, 17,
  261

\bibitem[{Vorobyov \& Basu(2009)}]{Vorobyov_2009}
Vorobyov, E.~I. \& Basu, S. 2009, Monthly Notices of the Royal Astronomical
  Society, 393, 822–837

\bibitem[{{Wada} {et~al.}(2011){Wada}, {Tanaka}, {Suyama}, {Kimura}, \&
  {Yamamoto}}]{Wada2011}
{Wada}, K., {Tanaka}, H., {Suyama}, T., {Kimura}, H., \& {Yamamoto}, T. 2011,
  \apj, 737, 36

\bibitem[{{Wardle}(2007)}]{2007Ap&SS.311...35W}
{Wardle}, M. 2007, \apss, 311, 35

\bibitem[{{Windmark} {et~al.}(2012){Windmark}, {Birnstiel}, {G{\"u}ttler},
  {Blum}, {Dullemond}, \& {Henning}}]{2012A&A...540A..73W}
{Windmark}, F., {Birnstiel}, T., {G{\"u}ttler}, C., {et~al.} 2012, \aap, 540,
  A73

\bibitem[{{Yang} {et~al.}(2018){Yang}, {Mac Low}, \&
  {Johansen}}]{2018ApJ...868...27Y}
{Yang}, C.-C., {Mac Low}, M.-M., \& {Johansen}, A. 2018, \apj, 868, 27

\bibitem[{{Yang} \& {Menou}(2010)}]{2010MNRAS.402.2436Y}
{Yang}, C.-C. \& {Menou}, K. 2010, \mnras, 402, 2436

\bibitem[{{Youdin} \& {Lithwick}(2007)}]{Youdin2007}
{Youdin}, A.~N. \& {Lithwick}, Y. 2007, \icarus, 192, 588

\bibitem[{{Zhao} {et~al.}(2016){Zhao}, {Caselli}, {Li}, {Krasnopolsky},
  {Shang}, \& {Nakamura}}]{2016MNRAS.460.2050Z}
{Zhao}, B., {Caselli}, P., {Li}, Z.-Y., {et~al.} 2016, \mnras, 460, 2050

\bibitem[{Zhu {et~al.}(2012)Zhu, Nelson, Dong, Espaillat, \&
  Hartmann}]{Zhu_2012}
Zhu, Z., Nelson, R.~P., Dong, R., Espaillat, C., \& Hartmann, L. 2012, The
  Astrophysical Journal, 755, 6

\end{thebibliography}


\onecolumn

\begin{appendix}

\section{Rayleigh adjustment for a nonuniform radial grid} \label{appendix:rayleigh adjustment}

\subsection{Context} \label{appendix:Context}

As explained in \citet{Delage2021}, a gas surface density profile resulting from steady-state accretion of our MRI-driven turbulence model (Sect.~\ref{sect:MRI-driven turbulence}) often leads to a striking mathematical discontinuity located at the dead zone outer edge. This discontinuity arises from the on/off criteria for active MRI, inducing a steep change in the local turbulent parameter $\alpha$ at each transition between the dead zone and the MRI-active region. As proposed by \citet{2010MNRAS.402.2436Y}, though, such a steep gas transition would not happen because the gas would not be stable and would rearrange itself due to turbulent diffusion on a dynamical timescale (Rayleigh adjustment process). In Model IV, we chose $\Sigma_{\rm{gas}}$ to follow the gas surface density profile obtained for the steady-state MRI-driven accretion solution corresponding to Model II, with MRN-like grain size distribution ($a_{\rm{min}} = 0.55 \, \mu$m, $a_{\rm{dist,Max}} = 1 \, \mu$m, and $p_{\rm{dist,Exp}} = -3.5$). Consequently, we need to apply Rayleigh adjustment in order to smooth the density gradient at the dead zone outer edge, and thus avoid any potential physically inconsistent state while running the dust evolution model employed. After applying Rayleigh adjustment, the new $\Sigma_{\rm{gas}}$ obtained is the gas surface density profile that we refer as "Steady-State 2" (SS2) in Table~\ref{tab:summary models}.

Below we revisit the simple algorithm put forward by \citet{2010MNRAS.402.2436Y} to implement Rayleigh adjustment. Particularly, we describe a new and more robust method that can be applied to any Rayleigh unstable gas surface densities, for a nonuniform radial grid.

\subsection{Method} \label{appendix:Method_Rayleigh_adjustment}

A radially one-dimensional diffusion equation in polar coordinates $(r, \phi)$ reads
\begin{ceqn}
\begin{equation} \label{E:diff}
\pder{\Sigma_{\rm{gas}}}{t} = \frac{1}{r}\pder{}{r}\left(r\mathcal{D}\pder{\Sigma_{\rm{gas}}}{r}\right),
\end{equation}
\end{ceqn}
where $\Sigma_{\rm{gas}}$ is the gas surface density and $\mathcal{D}$ is the diffusion coefficient. 
Eq.~\eqref{E:diff} represents a conservation law
\begin{ceqn}
\begin{equation} \label{E:claw}
\pder{\Sigma_{\rm{gas}}}{t} + \frac{1}{r}\pder{}{r}\left(r\mathcal{F}\right) = 0,
\end{equation}
\end{ceqn}
where the flux is defined by $\mathcal{F}(\Sigma_{\rm{gas}}(r,t); r, t) \equiv -\mathcal{D}\pder{\Sigma_{\rm{gas}}}{r}$.
Integrating Eq.~\eqref{E:claw} over a concentric ring from $r = r_1$ to $r = r_2$ and dividing the result by the area of the ring gives
\begin{ceqn}
\begin{equation} \label{E:intr}
\pder{}{t}\left(\frac{2}{r_2^2 - r_1^2}\int_{r_1}^{r_2}\Sigma_{\rm{gas}}\,r\mathrm{d}r\right) +
\frac{2}{r_2^2 - r_1^2}
\left[r_2\mathcal{F}(r_2,t) - r_1\mathcal{F}(r_1,t)\right] = 0.
\end{equation}
\end{ceqn}
Defining the cell average as
\begin{ceqn}
\begin{equation}
Q(t) \equiv \frac{2}{r_2^2 - r_1^2}\int_{r_1}^{r_2}\Sigma_{\rm{gas}}\,r\mathrm{d}r
\end{equation}
\end{ceqn}
and integrating Eq.~\eqref{E:intr} from $t = t_1$ to $t = t_2$ gives
\begin{ceqn}
\begin{equation} \label{E:intt}
Q(t_2) = Q(t_1) -
\frac{2\Delta t}{r_2^2 - r_1^2}
\left[r_2 F(r_2) - r_1 F(r_1)\right],
\end{equation}
\end{ceqn}

\noindent where $\Delta t \equiv t_2 - t_1$ and $F(r) \equiv \frac{1}{\Delta t}\int_{t_1}^{t_2} \mathcal{F}(r,t)\,\mathrm{d}t.$

Consider a nonuniform grid with a mapping from the index space $\rho$ to the physical space $r$, i.e., $r = r(\rho) \equiv r_\rho$. We adopt the convention that the cell edges are indexed by integers and the cell centers by half-integers.
Eq.~\eqref{E:intt} then gives the Godunov method for the $(n+1)$-th time step
\begin{ceqn}
\begin{equation} \label{E:god}
Q_{j+1/2}^{n+1} = Q_{j+1/2}^n -
\frac{2\Delta t}{r_{j+1}^2 - r_j^2}\left(r_{j+1}F_{j+1}^n - r_j F_j^n\right),
\end{equation}
\end{ceqn}
where
\begin{ceqn}
\begin{equation}
Q_{j+1/2}^n \equiv
\frac{2}{r_{j+1}^2 - r_j^2}\int_{r_j}^{r_{j+1}}\Sigma_{\rm{gas}}(r,t_n)\,r\mathrm{d}r,
\end{equation}
\end{ceqn}
and
\begin{ceqn}
\begin{equation} \label{E:flux}
F_j^n \equiv
\frac{1}{\Delta t}\int_{t_n}^{t_{n+1}} \mathcal{F}(r_j,t)\,\mathrm{d}t,
\end{equation}
\end{ceqn}
with $\Delta t \equiv t_{n+1} - t_n$.

To proceed, one could consider a Riemann problem and use the solution to evaluate Eq.~\eqref{E:flux}. Instead, we approximate it by assuming $\Sigma_{\rm{gas}}(r,t)$ is nearly unchanged over $t \in [t_n, t_{n+1}]$ and hence
\begin{ceqn}
\begin{equation}
F_j^n \approx
\mathcal{F}(r_j,t_n) =
\left.-\mathcal{D}\pder{\Sigma_{\rm{gas}}}{r}\right|_{r = r_j} =
\left.-\mathcal{D}\tder{\rho}{r}\pder{\Sigma_{\rm{gas}}}{\rho}\right|_{r = r_j}.
\end{equation}
\end{ceqn}
It can then be discretized using central differences:
\begin{ceqn}
\begin{equation}
F_j^n \approx -\mathcal{D}(r_j)\rho'(r_j)\left(Q_{j+1/2}^n - Q_{j-1/2}^n\right),
\end{equation}
\end{ceqn}
where $\rho(r)$ is the inverse function of $r(\rho)$. Therefore, Eq.~\eqref{E:god} becomes
\begin{ceqn}
\begin{equation}
Q_{j+1/2}^{n+1} = Q_{j+1/2}^n +
\frac{2\Delta t}{r_{j+1}^2 - r_j^2} \left[
    r_{j+1}\rho'(r_{j+1})\mathcal{D}(r_{j+1})\left(Q_{j+3/2}^n - Q_{j+1/2}^n\right) - r_j\rho'(r_j)\mathcal{D}(r_j)\left(Q_{j+1/2}^n - Q_{j-1/2}^n\right)
    \right].
\end{equation}
\end{ceqn}
If the diffusion coefficient $\mathcal{D}$ is a constant, it can be simplified as
\begin{ceqn}
\begin{equation} \label{E:update}
Q_{j+1/2}^{n+1} = Q_{j+1/2}^n +
\frac{2\mathcal{D}\Delta t}{r_{j+1}^2 - r_j^2} \left[
    r_{j+1}\rho'(r_{j+1})\left(Q_{j+3/2}^n - Q_{j+1/2}^n\right) -
    r_j\rho'(r_j)\left(Q_{j+1/2}^n - Q_{j-1/2}^n\right)\right].
\end{equation}
\end{ceqn}

Next, we find the stability condition for the time step $\Delta t$.
Assume that
\begin{ceqn}
\begin{equation}
Q_{j+1/2}^n \sim e^{ikj},
\end{equation}
\end{ceqn}
where $k$ is any wavenumber in the index space $\rho$. Substituting this into Eq.~\eqref{E:update}, a von Neumann stability analysis gives
\begin{ceqn}
\begin{align}
Q_{j+1/2}^{n+1} &\sim
e^{ikj}\left\{1 +
    \frac{2\mathcal{D}\Delta t}{r_{j+1}^2 - r_j^2} \left[
        r_{j+1}\rho'(r_{j+1})\left(e^{+ik} - 1\right) -
        r_j\rho'(r_j)\left(1 - e^{-ik}\right)\right]\right\}\\&=
e^{ikj}\left[1 +
    (\alpha + \beta)\left(e^{+ik} - 1\right) -
    (\alpha - \beta)\left(1 - e^{-ik}\right)\right]\nonumber\\&=
e^{ikj}[1 - 2\alpha(1 - \cos k) + 2i\beta\sin k] \equiv A e^{ikj},\nonumber
\end{align}
\end{ceqn}
where
\begin{ceqn}
\begin{equation}
\begin{aligned}
\alpha &\equiv \frac{\mathcal{D}\Delta t\left[
    r_{j+1}\rho'(r_{j+1}) + r_j\rho'(r_j)\right]}{r_{j+1}^2 - r_j^2},\\
\beta  &\equiv \frac{\mathcal{D}\Delta t\left[
    r_{j+1}\rho'(r_{j+1}) - r_j\rho'(r_j)\right]}{r_{j+1}^2 - r_j^2},
\end{aligned}
\end{equation}
\end{ceqn}
and
\begin{ceqn}
\begin{equation}
A \equiv 1 - 2\alpha(1 - \cos k) + 2i\beta\sin k.
\end{equation}
\end{ceqn}
We note that $\alpha > 0$ and $\beta < \alpha$ if $r(\rho)$ is a strictly monotonically increasing or decreasing function of $\rho$, which should be always the case. To be stable, $|A| = (1 - 2\alpha + 2\alpha\cos k)^2 + 4\beta^2\sin^2 k \le 1$. The extrema of $|A|$ occur at $\sin k = 0$ or $\cos k = \alpha(2\alpha - 1) / 2(\alpha^2 - \beta^2)$. The first leads to $(1 - 4\alpha)^2 \le 1$, or $0 \le \alpha \le 1/2$. The second results in $(\alpha - 2\beta^2)^2 \ge 0$, which is always true.
Therefore, the algorithm of Eq.~\eqref{E:update} can be optimized by adopting
\begin{ceqn}
\begin{equation} \label{E:optimization}
\mathcal{D}\Delta t \equiv
\left\{2\max_j\left[\frac{r_{j+1}\rho'(r_{j+1}) + r_j\rho'(r_j)}{r_{j+1}^2 - r_j^2}\right]\right\}^{-1}.
\end{equation}
\end{ceqn}

To conclude, a general outline of the Rayleigh adjustment procedure would be as follows:
\begin{enumerate}
\item Evaluate the stability of the gas surface density profile using the stability criterion derived in \citet{2010MNRAS.402.2436Y} (either their Eq.~(3) for a simple one, or a generalization shown in their Sect.~4). In the present paper, we chose to follow their Eq.~(3).

\item If the gas surface density profile is unstable (which is the case in the steady-state accretion of our MRI-driven turbulence model), use the unstable profile as the initial condition and iterate Eq.~\eqref{E:update} with $\mathcal{D}\Delta t$ following Eq.~\eqref{E:optimization} until the profile has relaxed to marginal stability.

\item Use the relaxed gas surface density profile to run the 1D dust evolution model (similar to what is done in this paper), or to resume the next time-step of the 1D gas and dust evolution model (in this case Rayleigh adjustment needs to be applied at each time-step when solving for the gas).
\end{enumerate}

\section{Dependence of the steady-state MRI-driven accretion on the power-law dust size distribution} \label{appendix:Follow-up on the effect of a dust distribution on the steady-state MRI-driven accretion regime}

In this appendix we perform a follow-up on the effect of a fixed power-law dust size distribution on the steady-state MRI-driven accretion. Particularly, we investigate the impact of a variation in the dust distribution minimum grain size (Appendix~\ref{appendix:variation in the distribution minimum grain size}) as well as its exponent (Appendix~\ref{appendix:variation in the distribution exponent}).

\subsection{Variation in the distribution minimum grain size} \label{appendix:variation in the distribution minimum grain size}

\begin{figure*}
\centering
\includegraphics[width=\textwidth]{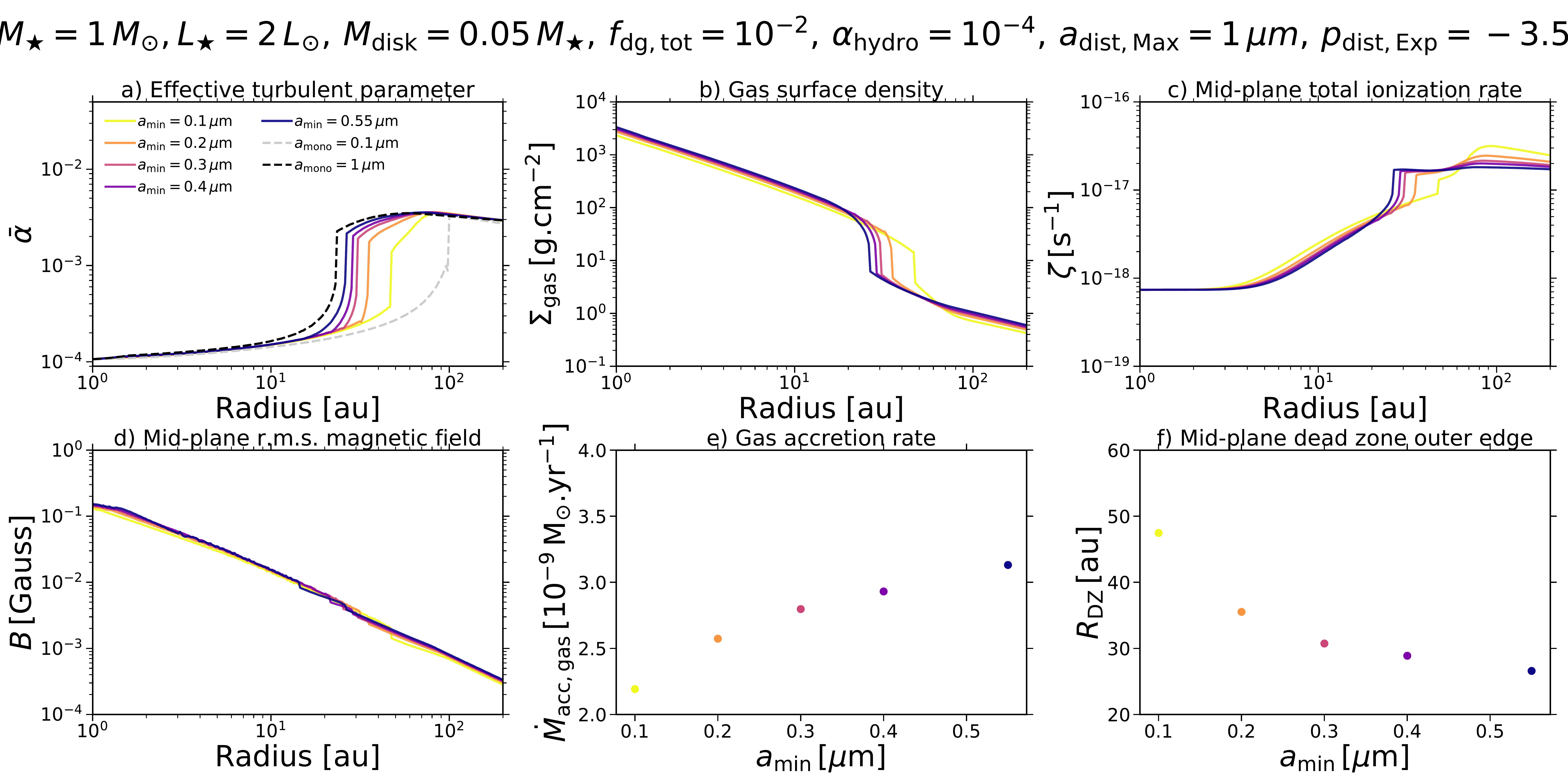}
 \caption{Impact on the steady-state MRI-driven accretion when varying the minimum grain size, $a_{\rm{min}}$, of the fixed power-law dust size distribution (Model II). Going from solid light-colored to dark-colored lines, $a_{\rm{min}}$ spans the range from $0.1 \, \mu$m to $0.55 \, \mu$m. \textit{Panels a--f} show the same quantities as in Fig.~\ref{fig:Vary_a_distMAX}, but for the model parameters $M_{\star} = 1 \, M_{\odot}$, $L_{\star} = 2 \, L_{\odot}$, $M_{\rm{disk}} = 0.05 \,M_\star$, $f_{\rm{dg,tot}} = 10^{-2}$, $\alpha_{\rm{hydro}} = 10^{-4}$, $a_{\rm{dist,Max}} = 1 \, \mu$m, and $p_{\rm{dist,Exp}} = -3.5$. For comparison, the dashed gray and black lines in \textit{Panel a} now display the steady-state quantity $\bar{\alpha}$ obtained assuming the limiting case of a mono-disperse population of dust with size $a_{\rm{mono}} = 0.1 \, \mu$m and $a_{\rm{mono}} = 1 \, \mu$m, respectively.}
 \label{fig:Vary_a_min}
\end{figure*}

We investigate the impact of a variation in $a_{\rm{min}}$, while fixing $a_{\rm{dist,Max}} = 1 \, \mu$m and $p_{\rm{dist,Exp}} = -3.5$. This set of simulations corresponds to Model II, and the results are presented in Fig.~\ref{fig:Vary_a_min}.

A higher $a_{\rm{min}}$ implies stronger MRI-driven turbulence overall, a higher gas accretion rate, and a more compact dead zone (Figs.~\ref{fig:Vary_a_min}a, \ref{fig:Vary_a_min}e and \ref{fig:Vary_a_min}f). Unlike what we saw in Sect.~\ref{sect:the effect of a dust distribution on the steady-state MRI-driven accretion regime}, though, we notice that a change in $a_{\rm{min}}$ primarily impacts the dead zone outer edge location, which can be almost twice closer from the central star for $a_{\rm{min}} = 0.55 \, \mu$m compared to $a_{\rm{min}} = 0.1 \, \mu$m. The gas accretion rate and the optimal r.m.s. magnetic field strength appear, indeed, to be weakly dependent on $a_{\rm{min}}$ (Figs.~\ref{fig:Vary_a_min}d and \ref{fig:Vary_a_min}e).

When a larger grain size is used for the dust distribution minimum size, the gas ionization degree becomes higher closer to the central star (Fig.~\ref{fig:Vary_a_min}c), since gas-phase recombination can dominate the recombination process from smaller radial distances. For increasing value of $a_{\rm{min}}$, the total grain surface area, $A_{\rm{tot}}$, decreases and the dust becomes less efficient in sweeping up free electrons and ions from the gas phase at smaller radial distances, hence the MRI being able to operate more easily. 

It is important to note that the distribution minimum grain size, $a_{\rm{min}}$, is a free-parameter in dust models with a distribution of sizes. From Fig.~\ref{fig:Vary_a_min}, it can be seen that $a_{\rm{min}}$ is crucial in determining the MRI-driven turbulence. We discuss the implications in Sect.~\ref{sect:Effect of the dust distribution minimum grain size}.  

\subsection{Variation in the power-law exponent} \label{appendix:variation in the distribution exponent}

\begin{figure*}
\centering
\includegraphics[width=\textwidth]{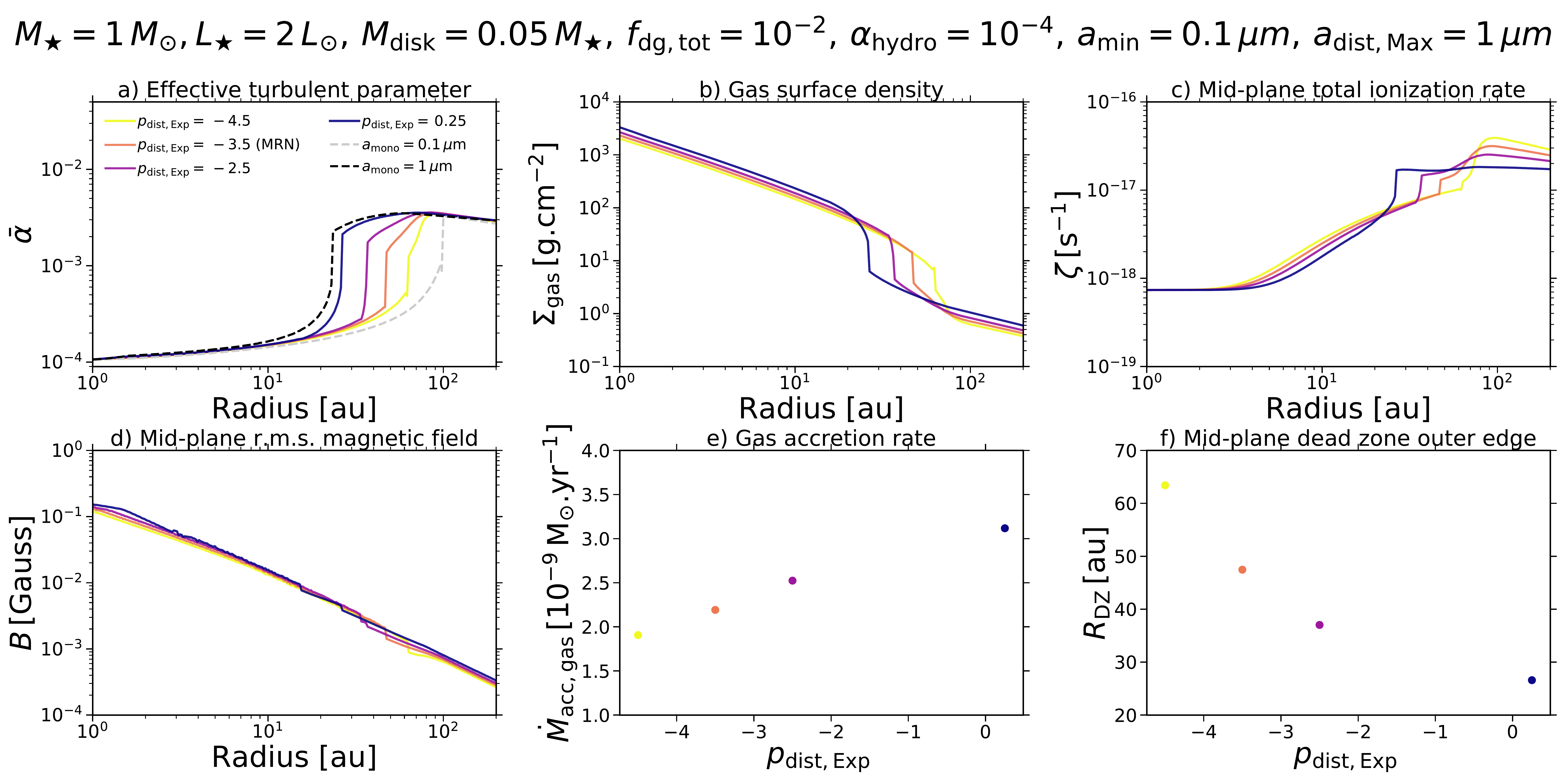}
 \caption{Impact on the steady-state MRI-driven accretion when varying the exponent, $p_{\rm{dist,Exp}}$, of the fixed power-law dust size distribution (Model III). Going from solid light-colored to dark-colored lines, $p_{\rm{dist,Exp}}$ spans the range from $-4.5$ to $0.25$. \textit{Panels a--f} show the same quantities as in Fig.~\ref{fig:Vary_a_distMAX}, but for the model parameters $M_{\star} = 1 \, M_{\odot}$, $L_{\star} = 2 \, L_{\odot}$, $M_{\rm{disk}} = 0.05 \,M_\star$, $f_{\rm{dg,tot}} = 10^{-2}$, $\alpha_{\rm{hydro}} = 10^{-4}$, $a_{\rm{min}} = 0.1 \, \mu$m, and $a_{\rm{dist,Max}} = 1 \, \mu$m. For comparison, the dashed gray and black lines in \textit{Panel a} now display the steady-state quantity $\bar{\alpha}$ obtained assuming the limiting case of a mono-disperse population of dust with size $a_{\rm{mono}} = 0.1 \, \mu$m and $a_{\rm{mono}} = 1 \, \mu$m, respectively.}
 \label{fig:Vary_p_distExp_a_min1e-5}
\end{figure*}

We investigate the impact of a variation in $p_{\rm{dist,Exp}}$, while fixing $a_{\rm{min}} = 0.1 \, \mu$m and $a_{\rm{dist,Max}} = 1 \, \mu$m. This set of simulations corresponds to Model III, and the results are presented in Fig.~\ref{fig:Vary_p_distExp_a_min1e-5}.

The power-law exponent ($p_{\rm{dist,Exp}}$) controls the relative proportion of the smaller and larger particles in the grain size distribution: A smaller or more negative value means that the dust distribution is skewed toward the smaller sizes, whereas a larger or more positive value means that it is skewed toward the larger sizes. Looking at Figs.~\ref{fig:Vary_p_distExp_a_min1e-5}a, \ref{fig:Vary_p_distExp_a_min1e-5}e, \ref{fig:Vary_p_distExp_a_min1e-5}f, a higher $p_{\rm{dist,Exp}}$ results in stronger MRI-driven turbulence overall, a higher gas accretion rate, and a more compact dead zone. In the same spirit as the previous section, we notice that $p_{\rm{dist,Exp}}$ mainly changes the dead zone outer edge (although the gas accretion rate and the optimal r.m.s. magnetic field strength vary more with $p_{\rm{dist,Exp}}$ than $a_{\rm{min}}$). The dead zone outer edge is, indeed, located at $\sim 63 \,$au for $p_{\rm{dist, Exp}} = -4.5$, whereas it is located at $\sim 27 \,$au for $p_{\rm{dist, Exp}} = 0.25$.

The gas ionization degree becomes higher closer to the central star when the dust size distribution is skewed toward the larger sizes (Fig.~\ref{fig:Vary_p_distExp_a_min1e-5}c). For increasing value of $p_{\rm{dist,Exp}}$, free electrons and ions are less likely to encounter, per unit volume and on average, the smaller particles of the dust size distribution. Instead, they primarily interact with the larger ones. The total grain surface area ($A_{\rm{tot}}$) thus decreases, allowing for the gas-phase recombination to dominate closer to the central star, hence the MRI being able to operate more easily.

Since the ionization chemistry is primarily controlled by the smaller sizes of the dust distribution, the MRI activity crucially depends on their relative proportion. Particularly, our results suggest that the MRI activity substantially increases in the regions of the disk where the dust size distribution is cut-off at micron-sized particles rather than submicron-sized particles. This can happen when submicron-sized particles get depleted enough due to effective grain coagulation and less frequent fragmentation.

\section{Temporal evolution of the dust surface density distribution for Models V and VI} \label{appendix:temporal evolution of the dust surface density distribution for Models V and VI}

We define the vertically integrated dust surface density distribution per logarithmic bin of grain size, $\sigma$, as \citep[][]{Birnstiel2010}
\begin{ceqn}
\begin{equation}
    \sigma(r,a) = \int_{-\infty}^{+\infty} \: n'_{\rm{dust}}(r,z,a) \: m(a) \: a \: dz.
    \label{eq:sigma}
\end{equation}
\end{ceqn}

Defining $\sigma(r,a)$ as in Eq.~\eqref{eq:sigma} makes it a grid-independent dust density unlike the mass integrated over each numerical bin ($\Sigma_{\rm{dust}}(r,a)$). This way, all plots of $\sigma(r,a)$ are meaningful without the knowledge of the size grid that we used. Below we show the temporal evolution of $\sigma(r,a)$ for Models V and VI. The one for Model IV is shown in Fig.~\ref{fig:Steady_MRI_v_frag1e2_sigma_dustVSgrainVSradius}. 
\begin{figure*}
\centering
\includegraphics[width=\textwidth]{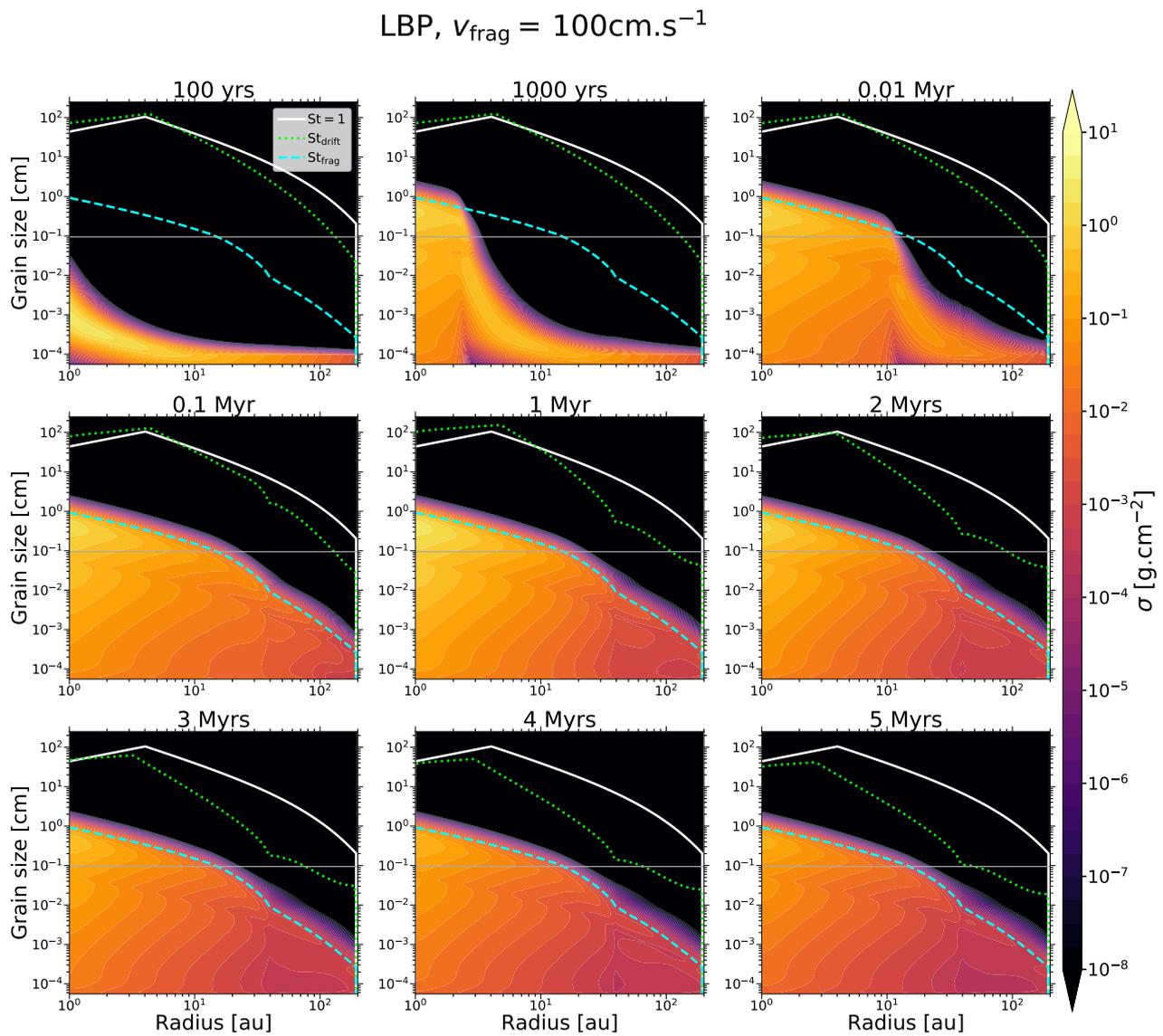}
  \caption{Same as in Fig.~\ref{fig:Steady_MRI_v_frag1e2_sigma_dustVSgrainVSradius}, except for Model V.}
 \label{fig:LBP_MRI_v_frag1e2_sigma_dustVSgrainVSradius}
\end{figure*}

\begin{figure*}
\centering
\includegraphics[width=\textwidth]{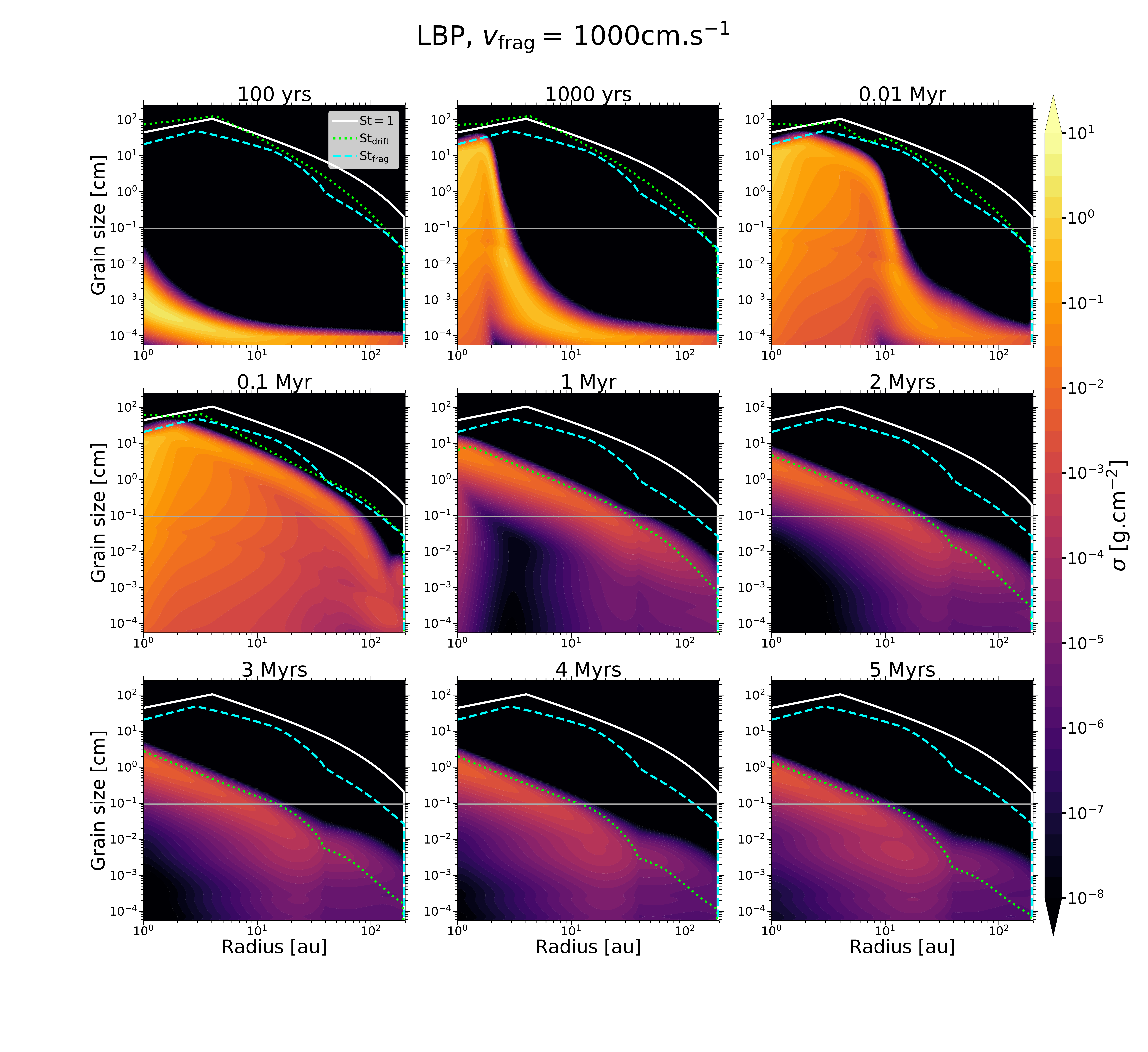}
  \caption{Same as in Fig.~\ref{fig:Steady_MRI_v_frag1e2_sigma_dustVSgrainVSradius}, except for Model VI.}
 \label{fig:LBP_MRI_v_frag1e3_sigma_dustVSgrainVSradius}
\end{figure*}

\end{appendix}

\end{document}